 \documentclass[aps,pra,letterpaper,superscriptaddress,twocolumn,showpacs,floatfix,10pt]{revtex4-1}

\usepackage{amsfonts}
\usepackage{amsmath}
\usepackage{amssymb}
\usepackage{graphicx}
\usepackage{epstopdf}
\usepackage{epsfig}
\usepackage{color}
\usepackage{mathtools}

\usepackage{hyperref}
\begin{document}

\title{Detecting and identifying 2D symmetry-protected topological, symmetry-breaking and intrinsic topological phases with modular matrices via tensor-network methods}

\author{Ching-Yu Huang}
 \affiliation{C. N. Yang Institute for Theoretical Physics and Department of Physics and Astronomy, State University of New York at Stony Brook, NY 11794-3840, United States}  

\author{Tzu-Chieh Wei}
 \affiliation{C. N. Yang Institute for Theoretical Physics and Department of Physics and Astronomy, State University of New York at Stony Brook, NY 11794-3840, United States}

\begin{abstract}
Symmetry-protected topological (SPT) phases exhibit nontrivial order if symmetry is respected but are adiabatically connected to the trivial product phase if symmetry is not respected. However, unlike the symmetry-breaking phase, there is no local order parameter for SPT phases. Here we employ a tensor-network method to compute the topological invariants characterized by the simulated modular $S$ and $T$ matrices to study  transitions in a few families of  two-dimensional (2D) wavefunctions which are $\mathbb{Z}_N$ ($N=2\, \&3$) symmetric. We find that in addition to the topologically ordered phases, the modular matrices can be used to identify nontrivial SPT phases and detect transitions between different SPT phases as well as between symmetric and symmetry-breaking phases. Therefore, modular matrices can be used to characterize various types of gapped phases in a unifying way. 
\end{abstract}

\date{\today}

\maketitle

\section{ introduction}
Landau symmetry breaking paradigm has been successful in explaining many observed phenomena from  ferromagnets to superconductors~\cite{WenBook}, exploiting notions such as long-range order and off-diagonal long-range order~\cite{Yang}. Such phases can be characterized by local order parameters, associated with certain symmetry and its breaking. However, it was recognized that systems such as fractional quantum Hall effect~\cite{TsuiStormerGossard,Laughlin} and quantum spin liquids~\cite{Kalmeyer,CSL,WenTO,ReadSachdev}
 cannot be described by local order parameters and they exhibit what we call topological orders (TO)~\cite{Wen_STmatrix_1989,Wen_Niu_STmatrix_1990,BravyiHastingsMichalakis10}. Classifying and characterizing them is an important task in understanding possible phases of matter in condensed matter physics, a pursuit no less fundamental than characterizing the periodic table of atoms.

Topological orders~\cite{Wen_STmatrix_1989,Wen_Niu_STmatrix_1990}, emergent phenomena usually in two spatial dimensions, appear in gapped systems displaying (i) ground-state degeneracy on multiconnected topology, (ii) local indistinguishability of degenerate ground states, (iii) fractional statistics of excitations, and (iv) the so-called long-range entanglement. Topological orders have also been much explored in the context of topological quantum 
computation, as properties such as (ii) and (iii), can give rise to natural  protection and robust processing of quantum information~\cite{Kitaev20032,TopolQC}. 
Much progress has since been made in understanding these more quantitatively. For example,  the long-range entanglement~\cite{Chen_Gu_Wen_2DQSRG_2010}
 in topological orders can be quantified by the so-called topological entanglement~\cite{Hamma,KitaevPreskill,LevinWenEE}, which is a deficit to the entanglement area law~\cite{AreaLaw}.
 The fractional statistics of excitations can be quantified by the so-called modular matrices $S$ and $T$, which are the (generically) non-Abelian geometric phases of the degenerate ground states on the two-dimensional torus~\cite{WenTO,Modular}.  They can be obtained via the wavefunction overlaps~\cite{He_Moradi_wen_GPTRG_2014,Moradi_wen_universalTO_2015}, conveniently using the basis of the so-called minimally entropy states  or minimally entangled states (MES)~\cite{Zhang_Ashvin_quasiparticle_2012,ZhuGongHaldaneSheng,Vidal_TOcylinder_2013}.

In addition to the above intrinsic topological orders, recently it was found that symmetry-protected topological (SPT) order can emerge if the ground state (which is unique) does not break global symmetry of the Hamiltonian, also in contrast to the spontaneous symmetry-breaking Landau paradigm. However, their excitations do not have fractional statistics in the bulk. Topological insulators and superconductors are such an prominent example~\cite{KaneMele,BernevigZhang,MooreBalents,FuKaneMele,Roy,Andreas_classification_TI_2008,Kitaev_TI_2009,Hasan_TI_2010,Qi_Zhang_TI_2011,Hasan_Moore_3DTI_2011,Wen_SPTfermion_2012}.
SPT phases have topological order that is not characterized by a local order parameter either, but their existence requires symmetry to be preserved~\cite{GuWen,ChenGuWen1D,Schuch_classifyMPS_2011,Fidkowski_Kitaev_1Dfermion_2011,Chen_Gu_Wen_classification1Dspin_2011,Chen_2DCZX_2011}. Ground states of non-trivial SPT phases cannot be continuously connected to trivial product states without either closing the gap or breaking the protecting symmetry.  Even though there is a unique ground state in the bulk,  for the open boundary condition, there are gapless excitations or nontrivial surface states at the boundary. 

In one dimension, the classification of SPT phases has been understood extensively~\cite{Pollmann_1DSPT_2010,ChenGuWen1D,Ari_Pollmann_1Dfermion_2011,Pollmann_Masaki_1DSPT_2012} and is considered complete~\cite{Chen_Gu_Wen_classification1Dspin_2011,Schuch_classifyMPS_2011,Fidkowski_Kitaev_1Dfermion_2011}. The characterization there can be obtained via the inequivalent projective representations corresponding to the on-site symmetry.  
In higher dimensions $d$  the use of $(d+1)$-th group cohomology gives rise to systematic construction  of large classes of  interacting bosonic  SPT phases~\cite{Chen_cohomology_science_2012,Chen_cohomology_2013}. 
In addition to cohomology, there have been alternative understanding of these SPT phases. In 2D, it was argued that 
any SPT phase characterized by third group cohomology $H^3[G,U(1)]$ is dual to a gauge theory of a finite group $G$~\cite{Levin_Gu_gaugeSPT_2012}.  Through this duality, it is possible to characterize 2D SPT phases using topological invariants such as modular $S$ and $T$ matrices and combinations of them~\cite{Hung_Wen_SPTinvariant_2014}.
In addition to cohomology and gauging, the classification of SPT phases can actually be understood via other approaches, such as (1) anomalous symmetry action at the boundary~\cite{ElseNayak}, (2) the nonlinear sigma model with a topological theta term~\cite{BiXu01,BiXu02}, and (3) gauge fields for symmetry twists~\cite{WangGuWen}, and (4) cobordism~\cite{Kapustin}, possibly leading to SPT phases beyond cohomology~\cite{VishwanathSenthil}.

 Given a Hamiltonian, how do we know whether there are any intrinsic or symmetry-protected topological orders in the phase diagram and how to detect them? Detecting and identifying these topological orders is also of practical importance as the classification only tells us what phases are possible in principle. This is a challenging task in general. For intrinsic TO, methods such as exact diagonalization~\cite{ZhuGongHaldaneSheng,Siddhardh14}, quantum Monte Carlo~\cite{Melko,ZhangGrover}, density-matrix renormalization group (DMRG) calculations~\cite{YanHuseWhite,DepenbrockKagome,Jiang2012,Vidal_TOcylinder_2013,Zalete_TOFQHE_2013,He14_dmrg} and tensor networks~\cite{PoilblancRVB,SchuchTopological2013,OrusTopologicalGE,He_Moradi_wen_GPTRG_2014,Huang_Wei_Z3TO_2015} have been applied to extract the topological entanglement, MES, and modular matrices.
 For SPT phases, much success has been limited to one dimension~\cite{GuWen,PollmannTurner}. Recently, the so-called strange correlator (SC)~\cite{XuStrangeCorrelator} has been proposed by Xu and collaborators and used to detect nontrivial SPT phases in both one~\cite{Wierschem1D} and two  dimensions~\cite{XuStrangeCorrelator}. Using the so-called momentum polarization (MP) originally developed for toplogical order~\cite{MomentumPolarization}, Zaletel recently proposed to use (i) the statistical spin $s_g$ of the $g$-charge and $g$-flux composite, where $g$ is an element in an Abelian group $G$ and (ii) the projective representation for a $g$-flux as a characterization for 2D SPT phases with an Abelian symmetry group~\cite{Zaletel14}.

 Here we employ tensor-network (TN) methods~\cite{VerstraeteMurgCirac,OrusTN} to obtain the topological invariants, i.e., the modular matrices
  $S$ and $T$ for detecting and identifying 2D topological orders, both intrinsic and symmetry-protected ones. It was previously shown by He, Moradi and Wen~\cite{He_Moradi_wen_GPTRG_2014} how to do this for topological phases via wavefunction overlaps~\cite{Moradi_wen_universalTO_2015}. Our main contribution here is to show and demonstrate how $S$ and $T$ can be obtained for  2D SPT phases, based on a theoretical proposal by Hung and Wen~\cite{Hung_Wen_SPTinvariant_2014}. We also find that $S$ and $T$ can be used to characterize symmetry-breaking phases as well, which have been studied recently using tensor-network methods on infinite cylinders~\cite{Norbert_SB_PEPS_2015}. For convenience we shall refer to our method as the tensor-network scheme for modular $S$ and $T$ matrices (tnST). As will be demonstrated below, our tnST scheme also applies to topological order. Therefore, modular matrices provide a potential unifying picture and tool for detecting and characterizing gapped phases.
A full-fledge application would involve (i) solving the Hamiltonian and obtaining the ground state wavefunction(s)~\cite{VerstraeteMurgCirac,OrusTN}, (ii) identifying the symmetry and the associated matrix-product operator (MPO)~\cite{VerstraeteTO_MPO_2014,VerstraeteSPT_MPO_2014}, and then (iii) employing our approach here to obtain the modular matrices.  To demonstrate the applicability of our method, we shall test it on a few families of wavefunctions that are either topologically ordered or symmetric under certain group action  (whose parent Hamiltonians can be constructed in principle; see Appendices~\ref{sec:firstZ2} and~\ref{sec:secondZ2} for two families of $\mathbb{Z}_2$ SPT Hamiltonians), bypassing the step of obtaining the unknown ground-state wavefunction from a given Hamiltonian and the associated MPO's.

 Classification of SPT phases~\cite{Chen_Gu_Wen_classification1Dspin_2011,Schuch_classifyMPS_2011,Fidkowski_Kitaev_1Dfermion_2011} was facilitated by the use of the TN states, also known as projected entangled pair states (PEPS) or tensor-product  states (TPS)~\cite{VerstraeteMurgCirac,OrusTN},  which were motivated from  generalization of the DMRG invented by White~\cite{DMRG} to higher dimensions. They can be regarded as variational wavefunctions. The matrix product states (MPS) in 1D turns out to be the form of wavefunction that the DMRG algorithm converges to~\cite{OstlundRommer,Perez-GarciaVerstraeteWolfCirac07}.  The TN states have also been shown to exactly
represent a large class of topological states, including both
nonchiral~\cite{TNstringnet,TNstringnet2,RVBPEPS,tcmera,KonigReichardtVidal09,VerstraeteTO_MPO_2014} and chiral~\cite{chiralPEPS1,chiralPEPS2,chiralPEPS3,YangChiral} topological order.
The evaluation of observables with MPS is efficient, whereas that with TN states in two or higher dimensions is a hard problem for exact contraction. But approximation schemes have been developed so that accuracy can be improved by increasing the size of tensors~\cite{VerstraeteMurgCirac,OrusTN}, with schemes including very recent development of powerful methods such as the tensor network renormalization (TNR)~\cite{TNR1,TNR2}. The specific scheme that we shall adopt for evaluating modular matrices via wavefunction overlaps is the so-called higher order tensor renormalization group (HOTRG) developed by Xiang and collaborators~\cite{Xiang_HOTRG_2012}. But other contraction schemes can be used as well.

Even though the models that we study in this paper are not realized by any natural materials, they are useful for advancing our  understanding of how to develop theoretical and numerical tools for probing their topological properties and provide a direct comparison of our numerical results with theoretical analysis. On the other hand, these models may be engineered in the laboratory. Materials have been discovered, such as ${\rm CsNiCl}_3$ and ${\rm RbNiCl}_3$ for realizing Haldane chains~\cite{EndState} (which is an exemplary SPT phase) and frustrated magnets such as herbertsmithite for realizing spin liquid~\cite{Balents}, and these systems are well approximated by lattice or spin Hamiltonians. Moreover, recent rapid development in quantum simulations using cold atoms, Rydberg atoms, trapped ions, cavities, photonics, and superconducting qubits, etc. have been proposed and developed to emulate lattice Hamiltonians, and possible exotic Hamiltonians (such as intrinsic and symmetry-protected topological orders) not necessarily existing naturally~\cite{ColdAtomDuan,Exchange,OptLatt,Ryberg,Greiner,Photonic,Cavities,Monroe,SC,Honeycomb}. It is possible that models related (not necessarily identical) to ones studied here  might be realized in some of these quantum simulation playgrounds.

Let us summarize our results and the structure of the remaining paper. In Sec.~\ref{sec:TheoryST} we briefly review the consideration of modular matrices and how they can be evaluated in both intrinsic and symmetry-protected topological phases.  We also present an analytic calculation of the modular matrices for fixed-point wavefunctions and symmetry action defined by 3-cocycles of a group $G$ in Ref.~\cite{Chen_cohomology_2013}. This gives us a basis to compare with our numerical results presented below.  In Sec.~\ref{sec:HOTRG_ST} we show how to use the real-space renormalization procedure of HOTRG to evaluate the modular matrices. In Sec.~\ref{sec:Numerical} we present results on various families of wavefunctions, for which parent Hamiltonians can be constructed. First we consider a one-parameter family of wavefunctions that interpolate between a $\mathbb{Z}_2$ toric-code topological phase and a trivial product phase, as well as one that is between a $\mathbb{Z}_2$ double-semion topological phase and a trivial product phase. Both were considered previously by He et al.~\cite{He_Moradi_wen_GPTRG_2014} by a gauge-symmetry-preserving tensor renormalization (GSPTRG)  approach. We use these wavefunctions to benchmark our method. Then we consider the so-called Ising PEPS~\cite{VerstraeteIsingPEPS,Norbert_SB_PEPS_2015}, which by construction has a quantum phase transition that corresponds to the classical Ising transition, and demonstrate that our modular-matrix approach can accurately detect such a transition between symmetric and spontaneous symmetry-breaking phases. For $\mathbb{Z}_N$ symmetry, there are one trivial and $(N-1)$ nontrivial SPT phases. We then construct families of wavefunctions that aim to interpolate from one to another symmetric phases for $N=2\&3$. We find that the  these SPT phases are  separated by a spontaneous symmetry-breaking phase. Finally, for $\mathbb{Z}_2$ SPT phases, we are able to find one family of wavefunctions for which a direct transition  connects the two SPT phases, which appears to be continuous.  Outside the two SPT phases actually lies a symmetry-breaking phase at large values of the parameter. We conclude in Sec.~\ref{sec:conclusion}, discuss different scenarios that have been proposed regarding transitions between two SPT phases~\cite{Tsui2015330}, and point out possible future directions.

\begin{figure}
   \includegraphics[width=0.5\textwidth]{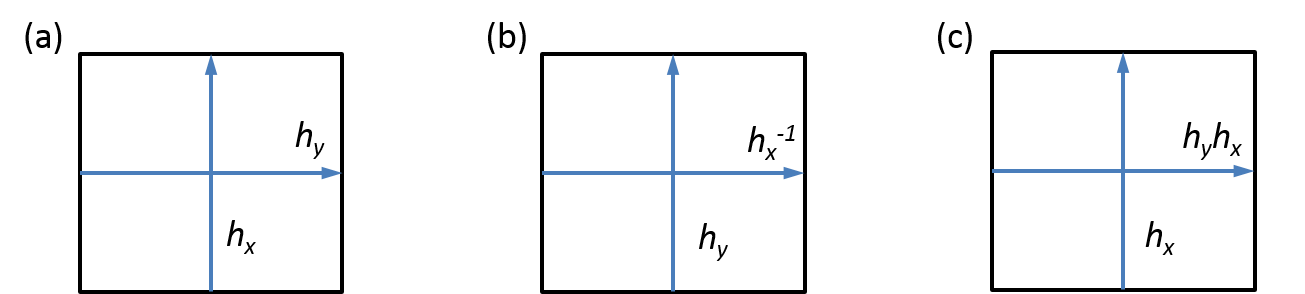}
  \caption{\label{fig:Modular}
  The effect of modular transformation. (a) Original string operators or symmetry twists indicated by $h_x$ and $h_y$. (b) After the modular $\hat{s}$ transformation. (c) After the modular $\hat{t}$ transformation.}
\end{figure}

\section{Modular matrices in intrinsic and symmetry-protected topological order} \label{sec:TheoryST}
\subsection{Modular transformation and modular matrices in intrinsic TO}
The modular transformation on a 2-torus is the group $SL(2,Z)$ which is generated by the two operations:
\begin{equation}
\hat{s}=\left(\begin{array}{cc}
0 & 1 \\
-1 & 0 \end{array}\right), \ \ \hat{t}=\left(\begin{array}{cc}
1 & 1 \\
0 & 1\end{array} \right),
\end{equation}
where $\hat{s}$ is a $90^\circ$ rotation and $\hat{t}$ is the so-called Dehn twist. For a basis of degenerate ground states $\{|\psi_a\rangle\}$ in a topological phase, we can consider the action of these two generators~\cite{Wen_STmatrix_1989,Wen_Niu_STmatrix_1990,Modular}:
\begin{align}
& \langle \psi_a | \hat{s} | \psi_b \rangle = e^{-\alpha_S A+\mathfrak{o} (1/A)} {S}_{ab}  \notag  \\
& \langle \psi_a | \hat{t} | \psi_b  \rangle= e^{-\alpha_T A+\mathfrak{o} (1/A)} {T}_{ab},
\end{align} 
where modular matrices ${S}$ and ${T}$ can be regarded as representations of these two generators in the basis of degenerate ground states $\{|\psi_i\rangle\}$, $A$ is the area of the torus and $\alpha_S$ and $\alpha_T$ are some positive constants independent of the ground states.
 For intrinsic TO, the number of degenerate ground states on the 2-torus equals that of the types of anyonic excitations, and the elements of ${S}$ and ${T}$ represent mutual and self-statistics of anyons~\cite{WenTO,Modular,LevinWen_stringnet_2005,Wen_STmatrix_2012}. One can use, e.g.,  the trace of these matrices, ${\rm tr}({S})$ and ${\rm tr}({T})$), to characterize topological phases and transitions
 
For example, consider the set of orthonormal degenerate ground states $|\Psi_{i,j}\rangle$ ($i,j=0,..,N-1$) of a topological phase at the fixed point, such as the quantum double models~\cite{Kitaev20032} or string-nets~\cite{LevinWen_stringnet_2005}. For a suitable choice of basis, there exists a set of closed-loop string operators $W_k^{[h]}$ and $W_k^{[v]}$ (parallel to horizontal and vertical circles, respectively; $k=0,..,N-1$ and $W_0=I$) such that $W_i^{[h]}W_j^{[v]}|\Psi_{0,0}\rangle=|\Psi_{i,j}\rangle$, where $W$'s commute with one another and the Hamiltonian. These operators can be schematically represented as the twist operators shown in Fig.~\ref{fig:Modular}(a). Instead of carrying out the modular transformations on the degenerate ground states, one can equivalently consider how the string operators transform under the modular transformations, as shown schematically in Fig.~\ref{fig:Modular}(b)(c). Take the toric code for example, depending on how one constructs $|\Psi_{0,0}\rangle$, these $W$'s can be a string of Pauli X or Z operators. The closed-loop string operators act on the physical degrees of freedom, but their effect on ground states can be equivalently induced by using the corresponding closed-loop matrix product operators that act on the virtual or bond degrees of freedom in the tensor-network description of the ground-state wavefunctions~\cite{VerstraeteTO_MPO_2014,Buerschaper2014}. 
For explicit calculations of $S$ and $T$ matrices in the string-net models, we refer the readers to Refs.~\cite{LevinWen_stringnet_2005,Liu_STmatrix_2013}.

\begin{figure}
   \includegraphics[width=0.4\textwidth]{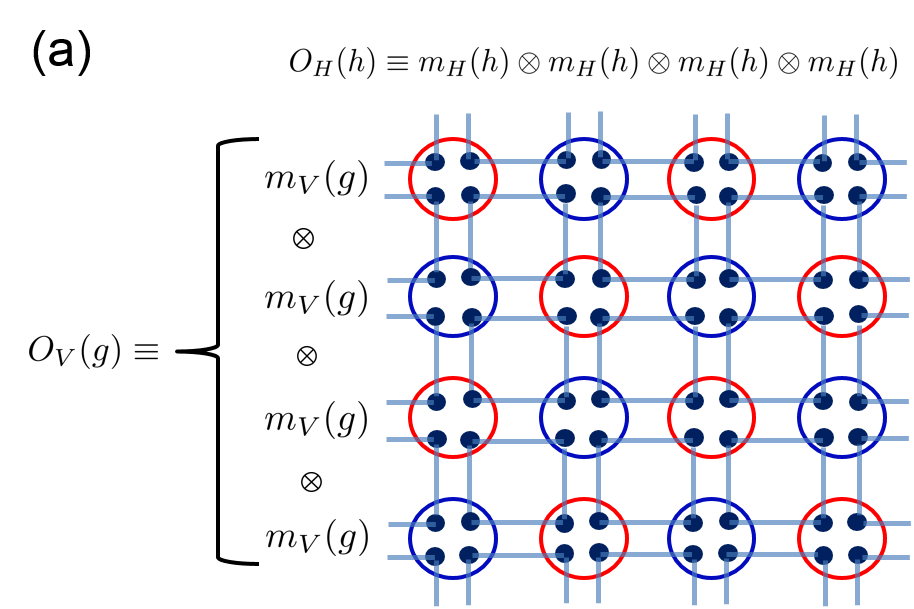}\vspace{0.5cm}
    \includegraphics[width=0.4\textwidth]{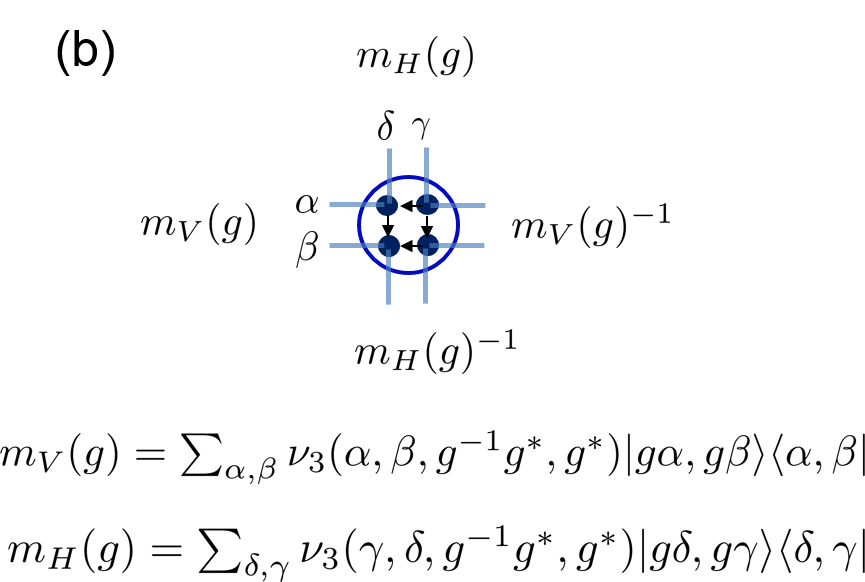}
  \caption{\label{fig:Twist}
 The symmetry twist operators $O_V(g)$ and $O_H(h)$. Each twist  can be visualized as an oriented branch cut, indicated by an arrow in the figure. The twist is in general described by a matrix-product operator. Here it is of a simple form, i.e., a product of local operators. The action of the local operator is shown in the bottom panel. The arrows inside a site (indicted by a circle) is the branching structure used in the original construction of the fixed-point wavefunction by the 3-cocyles or equivalently the symmetry operation in Ref.~\cite{Chen_cohomology_science_2012,Chen_cohomology_2013}. }
\end{figure}

\subsection{Modular S and T matrices for SPT phases} \label{sec:ST}

\begin{figure}[t]
   \includegraphics[width=0.5\textwidth]{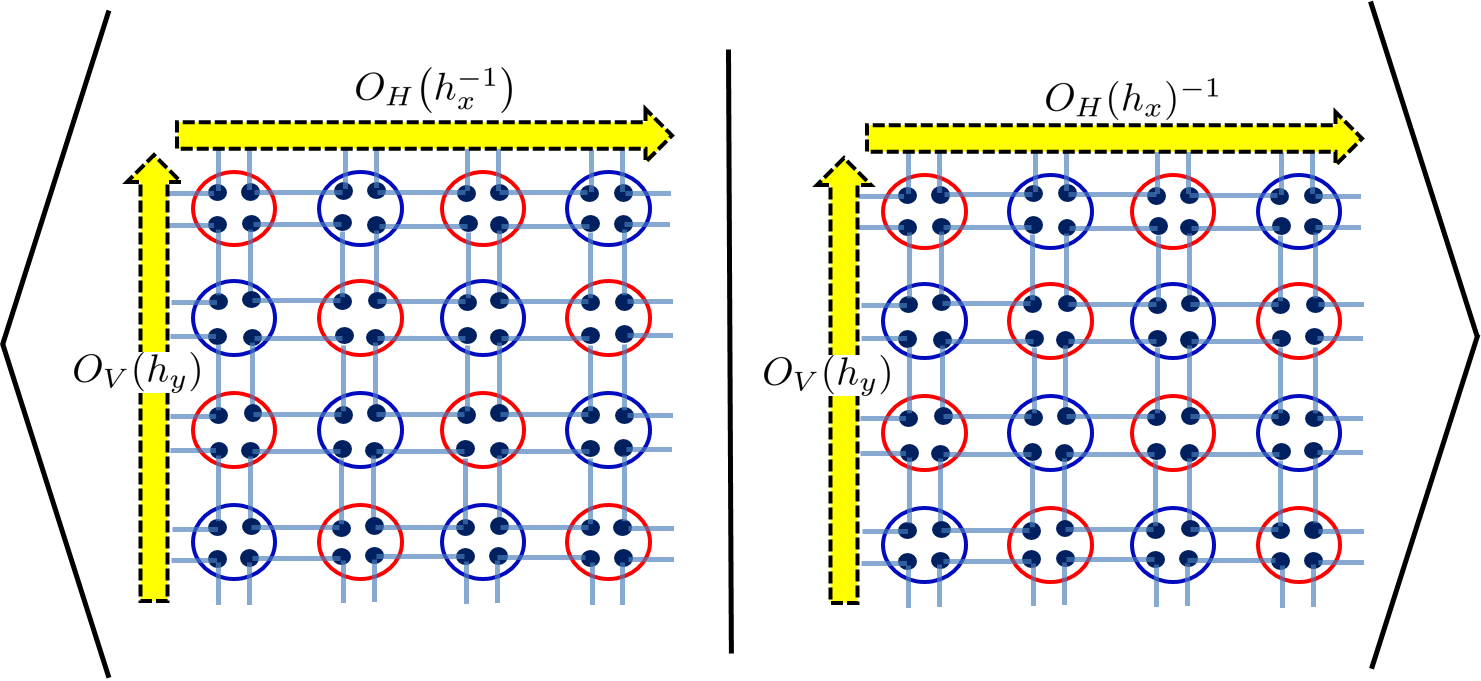}
  \caption{\label{fig:Smatrix}
  The modular $S$ matrix. $S_{(h_x',h_y'),(h_x,h_y)}=\langle \psi(h_x',h_y')|\psi(h_y,h_x^{-1})\rangle=\delta_{h_x',h_y}\delta_{h_y',h_x^{-1}}S_{(h_x,h_y)}
$. What is shown in the figure is the element $S_{(h_x,h_y)}$.}
\end{figure}

For SPT phases, the ground state $|\psi\rangle$ is unique. However,  one  can ``simulate'' the degenerate ground states by twisting the Hamiltonian and thus its twisted ground state~\cite{Hung_Wen_SPTinvariant_2014} (denoted by $|\psi(h_x,h_y)\rangle$) via a pair of commuting twist operators  $h_x$ and $h_y$ (of a group $G$ with $[h_x,h_y]=0$), as shown in Fig.~\ref{fig:Modular}(a). 
The lines labeled by $h_x$ and $h_y$ in Fig.~\ref{fig:Modular}(b)(c) for SPT phases represent the action of modular transformations on two such symmetry twists.  
Here we calculate $S$ and $T$ for the fixed-point SPT states with the symmetry action constructed by 3-cocycles of a group $G$ in Ref.~\cite{Chen_cohomology_2013}. In their construction, the fixed-point SPT states are schematically shown in Fig.~\ref{fig:Twist}a, where each site consists of four partons with degrees of freedom labelled by  $g\in G$ and these partons form GHZ-like entanglement with neighboring partons $\sum_{g}  |g_i=g,g_j=g,g_k=g,g_l=g\rangle/\sqrt{|G|}$, denoted by a square linking four such partons.  The symmetry action can be expressed in terms of local MPO, which acts on the local tensor as shown in Fig.~\ref{fig:Twist}(b).

\begin{figure}[t]
   \includegraphics[width=0.5\textwidth]{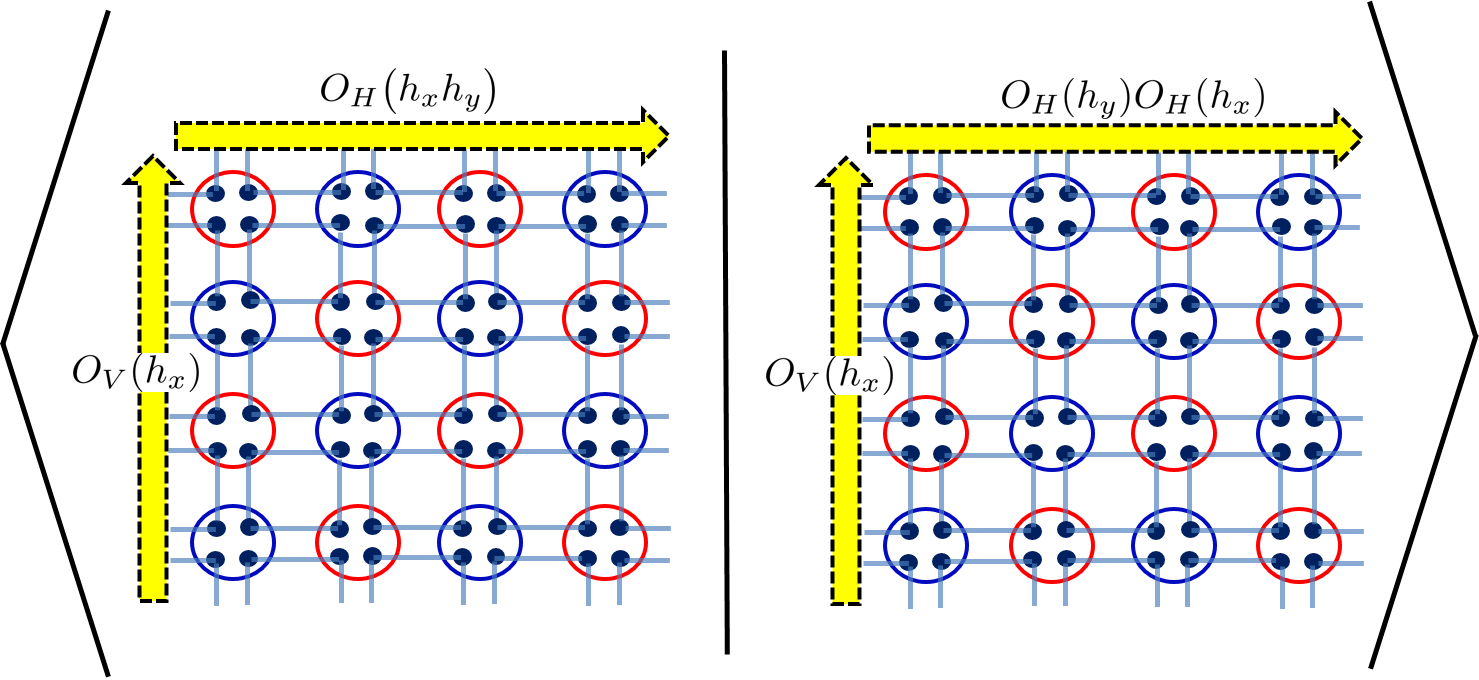}
  \caption{\label{fig:Tmatrix}
  The modular $T$ matrix. $  T_{(h_x',h_y'),(h_x,h_y)}=\langle \psi(h_x',h_y')|\psi(h_x,h_x h_y)\rangle=\delta_{h_x',h_x}\delta_{h_y',h_yh_x}T_{(h_x,h_y)}
$. What is shown in the figure is $T_{(h_x,h_y)}$.}
\end{figure}

Here we calculate the $S$ and $T$ according to the prescriptions  shown in Figs.~\ref{fig:Smatrix} and~\ref{fig:Tmatrix}. The elements of the modular matrices vanish unless the symmetry twists at the symmetry group level match: 
\begin{eqnarray}
S_{(h_x',h_y'),(h_x,h_y)}&\equiv&\langle \psi(h_x',h_y')|\hat{s}|\psi(h_x,h_y)\rangle\\
&=&\delta_{h_x',h_y}\delta_{h_y',h_x^{-1}}S_{(h_x,h_y)}.
\end{eqnarray}
\begin{eqnarray}
 T_{(h_x',h_y'),(h_x,h_y)}&\equiv&\langle \psi^{}(h_x',h_y')|\hat{t}|\psi^{}(h_x,h_y)\rangle \\
 &=&\delta_{h_x',h_x}\delta_{h_y',h_yh_x}T_{(h_x,h_y)}.
\end{eqnarray}
Using the symmetry twists defined in Fig.~\ref{fig:Twist} we obtain the modular matrices for the fixed-point wavefunction under the symmetry group $G$,
\begin{eqnarray}
S_{(h_x,h_y)}&=&\frac{1}{|G|}\sum_{a\in G}\frac{\nu_3(h_y a, g^*, h_x g^*,g^*)}{\nu_3(a,g^*,h_x g^*, g^*)},\\
T_{(h_x,h_y)}&=&\frac{1}{|G|}\sum_{a\in G}\frac{\nu_3\big( a,(h_yh_x)^{-1} g^*, h_x^{-1} g^*,g^*\big)}{\nu_3\big(h_x a,(h_y h_x)^{-1} g^*,h_x^{-1} g^*, g^*\big)},
\end{eqnarray}
where $\nu_3(g_0,g_1,g_2,g_3)$ is the 3-cocycle of the group $G$ and $g^*$ is an arbitrary but fixed element of $G$, which can be taken to be the identity element.
The exact forms are different from those in Hung and Wen~\cite{Hung_Wen_SPTinvariant_2014} constructed in an analogous way to the S and T matrices in the twisted quantum double models~\cite{Hu_twist_2013}. But they yield equivalent results for $\mathbb{Z}_N$ SPT phases. In particular, we have
\begin{eqnarray}
S_{(h_x,h_y)}&=&\delta_{h_x,e} + (1-\delta_{h_x,e}) e^{-\frac{2\pi k}{N}\bar{h}_y},\\
T_{(h_x,h_y)}&=& e^{i\frac{2\pi k}{N}\bar{h}_x\theta(\bar{h}_x+\bar{h}_y\ge N)},
\end{eqnarray}
where $e$ is the identity element, $\bar{g}$ represents the integer value (between $0$ and $N-1$) corresponding to the group element $g\in \mathbb{Z}_N$ and we have used the notation for the Heavside theta function $\theta(x\ge y)=1$ if $x\ge y$ and zero otherwise.

For example, the nonzero elements of S matrix for $\mathbb{Z}_2$ are
\begin{equation}
\left(
\begin{array}{ll}
\langle 0,0|\hat{S}|0,0\rangle=1 & \langle1,0|\hat{S}|0,1\rangle=1\\
\langle 0,1|\hat{S}|1,0\rangle=1 & \langle1,1|\hat{S}|1,1\rangle=(-1)^k 
\end{array}\right),
\end{equation}
and those for T matrix are
\begin{equation}
\left(
\begin{array}{ll}
\langle 0,0|\hat{T}|0,0\rangle=1 & \langle0,1|\hat{T}|0,1\rangle=1\\
\langle 1,1|\hat{T}|1,0\rangle=1 & \langle1,0|\hat{T}|1,1\rangle=(-1)^k 
\end{array}\right),
\end{equation}
where $k$ is used to indicate the inequivalent cohomology class (with $k=0$ being the trivial one).
As pointed out by Hung and Wen~\cite{Hung_Wen_SPTinvariant_2014}, due to the phase ambiguity the $\mathbb{Z}_N$ SPT phase is unambiguously characterized by the matrix elements of ${T}^N$ only, which we also have $\langle h,g |\hat{T}^N|h,g\rangle= \exp\big(2\pi i \bar{h}^2 k/N\big)$.

In the following, we shall be working with wavefunctions that are not at the fixed point and the overlaps of simulated degenerate ground states need to be evaluated numerically. We shall find that modular matrices can also be used to detect the transition between a symmetry-breaking and a symmetric phase, in addition to identifying the SPT phase. Including the application to intrinsic topological phases, the modular matrices provide a tool for characterizing 2D gapped phases.

Next, we describe how the modular matrices for topological order and symmetry protected topological order can be obtained via higher order tensor renormalization group method.

\section{ Modular matrices by higher order tensor renormalization group } \label{sec:HOTRG_ST}
Here we describe the contraction scheme that we shall use, i.e., the higher-order tensor renormalization group developed by Xiang and collaborators~\cite{Xiang_HOTRG_2012}. We then detail how modular matrices can be computed via wavefunction overlaps and how HOTRG can be applied in this context. But we remark that other contraction schemes can also be used.

\subsection{ Higher-order tensor renormalization group}

\begin{figure}[t]
\center{\epsfig{figure=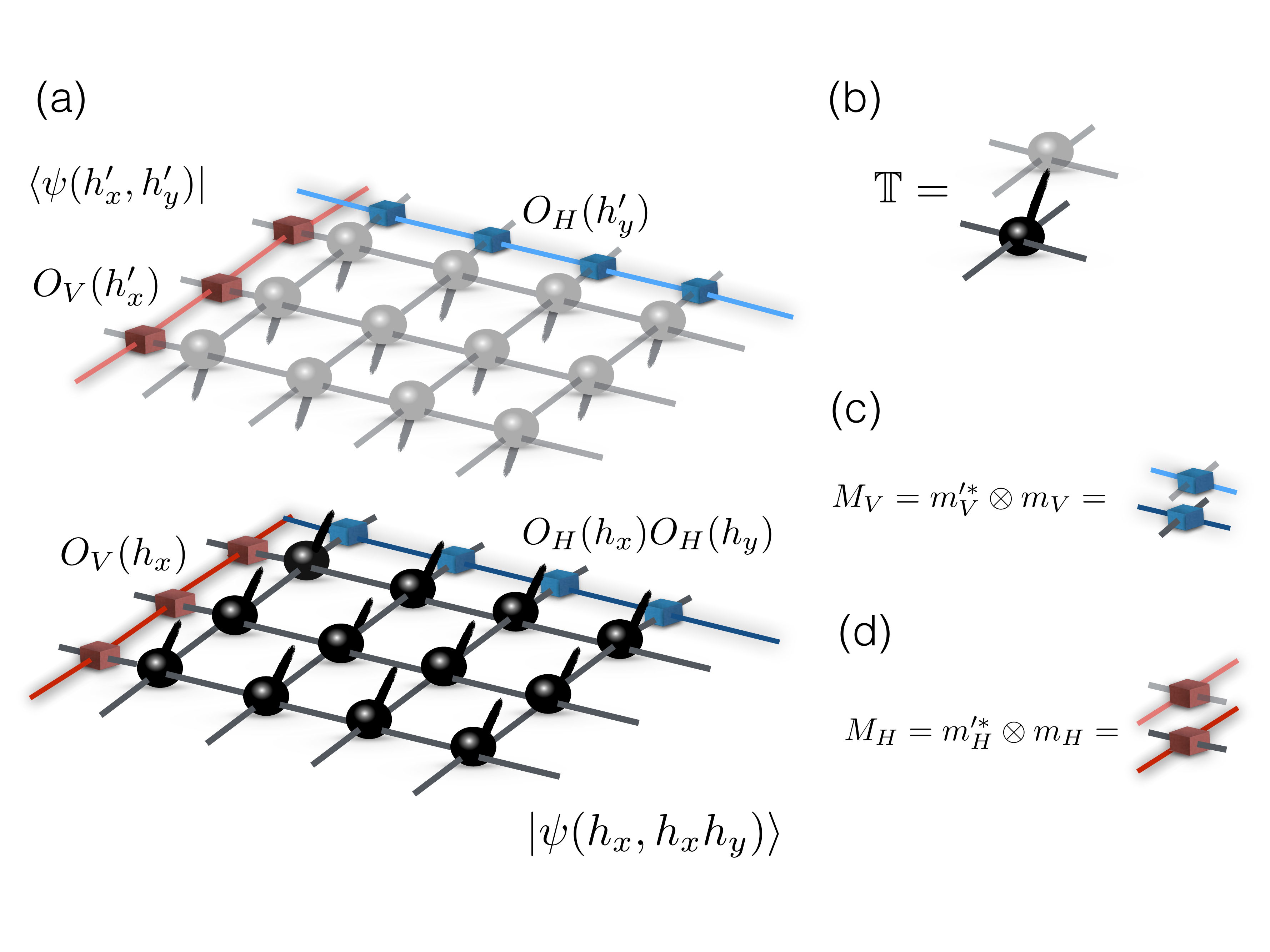,angle=0,width=9cm}}
\caption{(a)The tensor representation of overlap of the twist ground states. (b)The double tensor structure. (c) The double line structure of matrix product operators.  }
   \label{fig:Telements}
\end{figure}

A two-dimensional quantum state can be represented as a TPS, whose coefficients in a fixed basis are expressed as a contraction of a TN (i.e., a tensor trace),
\begin{align}
|\psi\rangle=\sum_{s_1,s_2,... s_m...}\text{tTr}(A^{s_1}A^{s_2}...A^{s_m}...)|s_1 s_2... s_m...\rangle, 
\label{TPS}
\end{align} 
where $A^s_{\alpha,\beta,\gamma,...}$ is a local tensor with a physical index $s$ and internal or bond indices $\alpha,\beta,\gamma,...$ and $\text{tTr}$ denotes tensor contraction of all the connected inner indices according to the underlying lattice structure; see Fig.~\ref{fig:Telements}(a).   
The norm square of TPS  is given by 
\begin{align}
\langle \psi |  \psi \rangle =\text{tTr} (\mathbb{T}^1  \mathbb{T}^2  \mathbb{T}^3... \mathbb{T}^m...   ), 
\label {ST_TPS}
\end{align} 
where the local {\it double tensor\/} $\mathbb{T}^i$ can be formed by merging two layers, tensors $A$ and $A^*$, with only physical indices contracted as shown in Fig.~\ref{fig:Telements}(b),
\begin{equation}
\mathbb{T}\equiv \sum _s  (A^s_{\alpha,\beta, \gamma,\delta... })  \times  (A^s_{\alpha',\beta', \gamma',\delta'... }) ^*.
\end{equation}
However, it is in general computationally hard to calculate exactly the tensor trace (tTr), i.e., the contraction of the whole tensor network in two and higher dimensions.
Several approximation schemes have been proposed as solutions in this context, such as iPEPS~\cite{Vidal_iPEPS_2009} algorithm, the corner transfer matrix method (CTMRG)~\cite{Nishino_CTMRG_1997,OrusCTM}, tensor renormalization approach~\cite{Levin_TRG_2007,Xiang_TRG_2008,Wen_Gu_Levin_TRG_2008}, and tensor network renormalization (TNR)~\cite{TNR1,TNR2},
all of which tackle the contraction problem essentially by making trancation and thus scaling down the computational complexity to the polynomial level.

Here we mainly use the HOTRG method to deal with the contraction~\cite{Xiang_HOTRG_2012}; this approach is a real-space renormalization procedure, where at each step tensors at neighboring sites are merged or coarse-grained and the bond dimension is truncated according to the relevance of the eigenvalues in the singular-value decomposition of the old tensors.
This technique will be applied for calculating the norm, wavefunction overlaps and  expectation values. 
The basic ingredient of the algorithm is schematically presented in Fig.~\ref{fig:HOTRG}.
In the HOTRG  coarse-graining,  merging of tensors is applied, e.g.,  along $y$ and subsequently $x$ direction via isometric transformations, and thus four sites are merged into a single site, reducing the lattice size by a factor of four. 
For example, merging two rank-4 $\mathbb{T}$ tensors along $y$ axis we obtain a new rank-6 tensor  $\mathbb{T'}= \sum_{\beta}  \mathbb{T}_{\alpha,\beta,\gamma,\delta}    \mathbb{T}_{\alpha',\beta',\gamma',\beta}$, where the indices of $\mathbb{T}$ start on the right  and go around clockwise to the top.
This can be regarded as a rank-4 tensor by formally grouping the two indices (represented as two lines) on the right to one index, and similarly the two on the left to another one.  The bond dimension of tensor $\mathbb{T'}$ along these directions is the square of the original bond dimension of tensor $\mathbb{T}$. 
Applying an appropriate isometry $U$ truncates the size of these squared bond dimensions to a fixed number, say, $D_{\rm cut}$,  and a truncated tensor   $\tilde{\mathbb{T}}$ can be obtained~\cite{Xiang_HOTRG_2012}. (The larger the $D_{\rm cut}$ the better accuracy can be obtained, but the larger memory and computational time will be required.)

To complete one RG step, two thus truncated $\mathbb{T'}$ tensors  are merged similarly along $x$ axis. 
Eventually, the entire network reduces to only a few sites and the tensor trace appearing in the norm square and physical observables of wave function can be calculated directly.  Equivalently, the effective lattice sites covered is exponential in the number of RG steps carried out.

\subsection{Modular matrices by wavefunction overlaps with symmetry twists using HOTRG}

In order to obtain the modular matrices, for topologically ordered phases, the degenerate ground states can be obtained by inserting the gauge transformation or closed string operators $(h_x,h_y)$ represented by matrix product operators (MPO) $ O_V(h_x)$ and $O_H(h_y)$ to virtual bond degrees of freedom. 
For the symmetry protected topological state $|\psi \rangle$,  $(h_x,h_y)$ represents the symmetry twists used  to ``simulate'' the degenerate ground states $|\psi (h_x,h_y)  \rangle$.
Once a basis of the simulated degenerate ground states are obtained, the modular matrices are calculated by  the matrix elements $\langle  \psi(h_x',h_y') | \hat{s} | \psi(h_x,h_y) \rangle$ and  $\langle  \psi(h_x',h_y') | \hat{t} | \psi(h_x,h_y) \rangle$, where the action of $\hat{s}$ and $\hat{t}$ on the symmetry twists $(h_x,h_y)$ as shown in Fig.~\ref{fig:Modular} can be used. 

In practice, to obtain the (simulated) degenerate ground states, we inset the matrix product operators $O_V(h_x)$ and $O_H(h_y)$ along the $y$ and $x$ directions, respectively, to the state $|\psi\rangle$, and the action of $\hat{s}$ and $\hat{t}$ will transform these MPO's; see  also Figs.~\ref{fig:Smatrix} and~\ref{fig:Tmatrix}.
Thus in carrying out the wavefunction overlaps, we also need to update the matrix product  operators as we update the local tensor in the procedure of HOTRG. 
After the coarse-graining procedure of the local tensor, the coarse-grained MPO's can be applied before the tensor contraction to determine the modular matrices. 
In brief, with an appropriate choice of MPO's, the same procedure works for both topologically ordered phases and symmetry-protected topological phases.

The modular matrices can be evaluated and monitored during the process of the tnST procedure (implemented using HOTRG), which can broken down to the following steps for convenience. 

(1) {\it Creating the basis set\/}. Given a ground state $|\psi\rangle$, thread  the symmetry twists for SPT phases (or gauge transformations for topologically ordered phases) $h_x$ and $h_y$ along $y$ and $x$ directions, respectively,  by applying the the  matrix product operators $O_V(h_x) = m_V\otimes m_V \otimes ... m_V $ and $O_H(h_y )= m_H\otimes m_H \otimes ... m_H $, respectively, and denote the resultant wavefunction as $|\psi(h_x,h_y)\rangle$; see Fig.~\ref{fig:Telements}(a).  This is the set of basis states that the modular transformations $\hat{s}$ and $\hat{t}$ will act on.

\smallskip (2) {\it Simulating the rotation and the Dehn twist\/}. 
As remarked earlier, the symmetry or string operators that are performed on the physical indices can be achieved by appropriate symmetry operations to internal indices. The rotation and Dehn twist can then be used to transform the symmetry twists (or string operators) themselves, schematically denoted as
\begin{align}
&   \big\langle  \psi(h_x',h_y') | \hat{t} | \psi(h_x,h_y)    \big\rangle =    \big\langle  \psi(h_x',h_y') | \psi(h_x,h_xh_y)       \big\rangle \notag \\
&   \big\langle  \psi(h_x',h_y') | \hat{s} | \psi(h_x,h_y)   \big\rangle =    \big\langle  \psi(h_x',h_y') | \psi(h_y,h_x^{-1})  \big\rangle \notag.
\end{align}
The tensor representation of  $ \big\langle  \psi(h_x',h_y') | \psi(h_x,h_xh_y) \big \rangle$ is shown in Fig.~\ref{fig:Telements}(a).

\begin{figure}[ht]
\center{\epsfig{figure=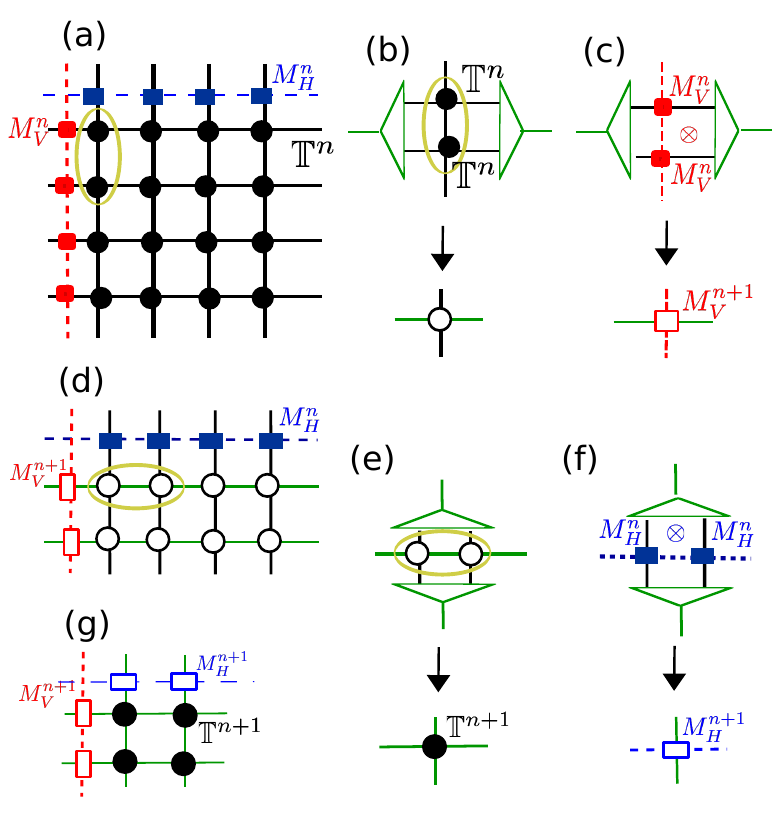,angle=0,width=9cm}}
\caption{ A HOTRG contraction of the tensor-network state along (a) y (d) x axis on the square lattice. Step of contraction and renormalization of two local tensor along (b) y-direction and (e) x-direction and renormalization of inner symmetry operators $M_H$ and $M_V$ in (c) (f). (g) At each step along $x$ and $y$ direction, four sites are contracted into a single site.}
   \label{fig:HOTRG}
\end{figure}

\smallskip (3) {\it Creating the double tensor and double MPO's\/}. Contract physical indices to from the double tensor $\mathbb{T}$ as shown in Fig.~\ref{fig:Telements}(b) and form the generalized doubled inner operators  $M_V = m'^{*}_V\otimes m_V$ and $M_H = m'^{*}_H\otimes m_H$ (see  Fig.~\ref{fig:Telements}(c)(d)), which  act on each  bond along vertical and horizontal twist lines, respectively, as shown in Fig.~\ref{fig:HOTRG}(a). 

When evaluating the modular  $T$ matrix: $\big\langle \psi(h_x',h_y')  | \psi(h_x,h_xh_y) \big\rangle $, the symmetry twists resulting from both the ket and the bra can be lumped into a generalized double-layer matrix product  operator defined by  $\mathbb{O}_H (h_x',h_x)= O_V(h_x')^{*} \otimes O_H(h_x) = M_H \otimes  M_H ... \otimes  M_H$ and $\mathbb{O}_V (h_y',h_xh_y)= O_V(h_y')^{*} \otimes O_V(h_x)O_V(h_y) = M_V \otimes  M_V ... \otimes  M_V$.  The generalized double-layer MPO for the modular $S$ matrix can be expressed similarly. It is these generalized MPO's that need to be coarse-grained as well.

% To illustrate this, let us take the  $\mathbb{Z}_2$ toric code as an example. In Eq.~(\ref{TOZ2_Tmatrix}),  the inner symmetry operator of the $\big\langle  \psi(\mathcal{Z},\mathcal{I})   | \psi(\mathcal{Z},\mathcal{ZZ})   \big\rangle$ is $\mathbb{O}_H(Z,Z) =  \mathcal{Z}^{*} \otimes \mathcal{Z} $ and  $\mathbb{O}_V(I,ZZ) =  \mathcal{I}^{*} \otimes \mathcal{Z}  \mathcal{Z} $. 
%For  $S$ matrix  $ \big\langle  \psi(h_x',h_y') | \psi(h_y,h_x^{-1})  \big\rangle \notag$,  for example, 
%the inner symmetry operator of the $\big\langle  \psi(\mathcal{Z},\mathcal{I})   | \psi(\mathcal{Z},\mathcal{Z}^{-1})   \big\rangle$ is $\mathbb{O}_H(Z,Z)  =  \mathcal{Z}^{*} \otimes \mathcal{Z} $ and  $\mathbb{O}_V(I,Z^{-1}) =  \mathcal{I}^{*} \otimes \mathcal{Z}^{-1} $. 

\smallskip (4) {\it Coarse-graining\/}. To coarse grain the local tensors and generalized MPO's, we contract the lattice along the  vertical and  then horizontal  directions. This scheme of coarse graining is shown in Fig.~\ref{fig:HOTRG}(a)(d)(g). 
First, two sites are contracted into single site by applying the isometric operators along the vertical direction as shown in Fig.~\ref{fig:HOTRG}(b). 
The same isometric operator  can also be used to coarse grain the  generalized inner symmetry operators $M_V$ in Fig.~\ref{fig:HOTRG}(c). 
Then, we coarse grain similarly two sites and  generalized inner symmetry operators $M_H$ along the horizontal direction as shown in  ~\ref{fig:HOTRG} (e)(f).
At each RG step, four sites are contracted into one, and the number of sites reduces by a factor of four. 

\smallskip (5) {\it Taking trace for overlaps and modular matrices\/}. Finally, we insert the matrix product operators $M_H^{f}$ and $M_V^{f}$ into the fixed-point double tensor $\mathbb{T}^{f}$ and take the trace of all internal indices in the fixed-point tensor (or  the tensor after sufficient number of RG steps) together with the coarse-grained generalized inner symmetry MPO's to determine the each element of the  modular matrices. We note that each element is normalized by the wavefunction norm square.

\smallskip
Let us remark that essentially the analytic calculations of $S$ and $T$ for the fixed-point SPT wavefunctions in Sec.~\ref{sec:ST} follow this line of thoughts, except that there is no need to do coarse-graining as the overlaps there can be done exactly. The results there also provide a basis for the numerics to compare with, especially away from fixed-point wavefunctions.

\section{Model constructions and numerical results} \label{sec:Numerical}

We will use several examples to demonstrate how to characterize and identity two-dimensional gapped quantum phases by using the modular matrices.
First, we demonstrate the usefulness of our tnST approach by applying to two models that exhibit $\mathbb{Z}_2$ topological order: the deformed toric-code and double-semion models in Sec.~\ref{subsec:TO}. These were previously studied in Ref.~\cite{He_Moradi_wen_GPTRG_2014} using GSPTRG. The purpose of examining these here is to illustrate that our tnST approach based on HOTRG works without imposing the gauge symmetry during the coarse-graining procedure.  
We then move on, in Sec.~\ref{subsec:SB}, to study symmetric and symmetry-breaking phases and illustrate this with the Ising PEPS model. We explain how to insert $\mathbb{Z}_2$ symmetry fluxes through symmetry twists to detect the transition point and characterize the two phases. This demonstrates that the modular matrices are also useful for identifying symmetry-breaking phases and long-range order.
In Sec.~\ref{subsec:ZNSPT} we construct models of  $\mathbb{Z}_N$ SPT states that are deformed from the group-cohomology fixed-point wavefunctions such that they remain symmetric under simple symmetry action of $\mathbb{Z}_N$ on physical spins through out all deformation. In particular, we study $\mathbb{Z}_2$ and $\mathbb{Z}_3$ cases, where different SPT phases are separated by a symmetry-breaking phase with long-range order through continuous quantum phase transitions.    
In Sec.~\ref{subsec:twoSPT} we construct another $\mathbb{Z}_2$ model that shows a direct continuous transition from one SPT phase (the trivial one)   to a nontrivial SPT phase.

\subsection{Topologically ordered phases}
\label{subsec:TO}

To test the validity of our tnST approach, we apply it to two topological models previously studied by He, Moradi and Wen using the gauge-symmetry-preserved tensor renormalization method (GSPTRG). We demonstrate that with our scheme using HOTRG without preserving the gauge symmetry, accurate results can be obtained.  

\begin{figure}[t]
\center{\epsfig{figure=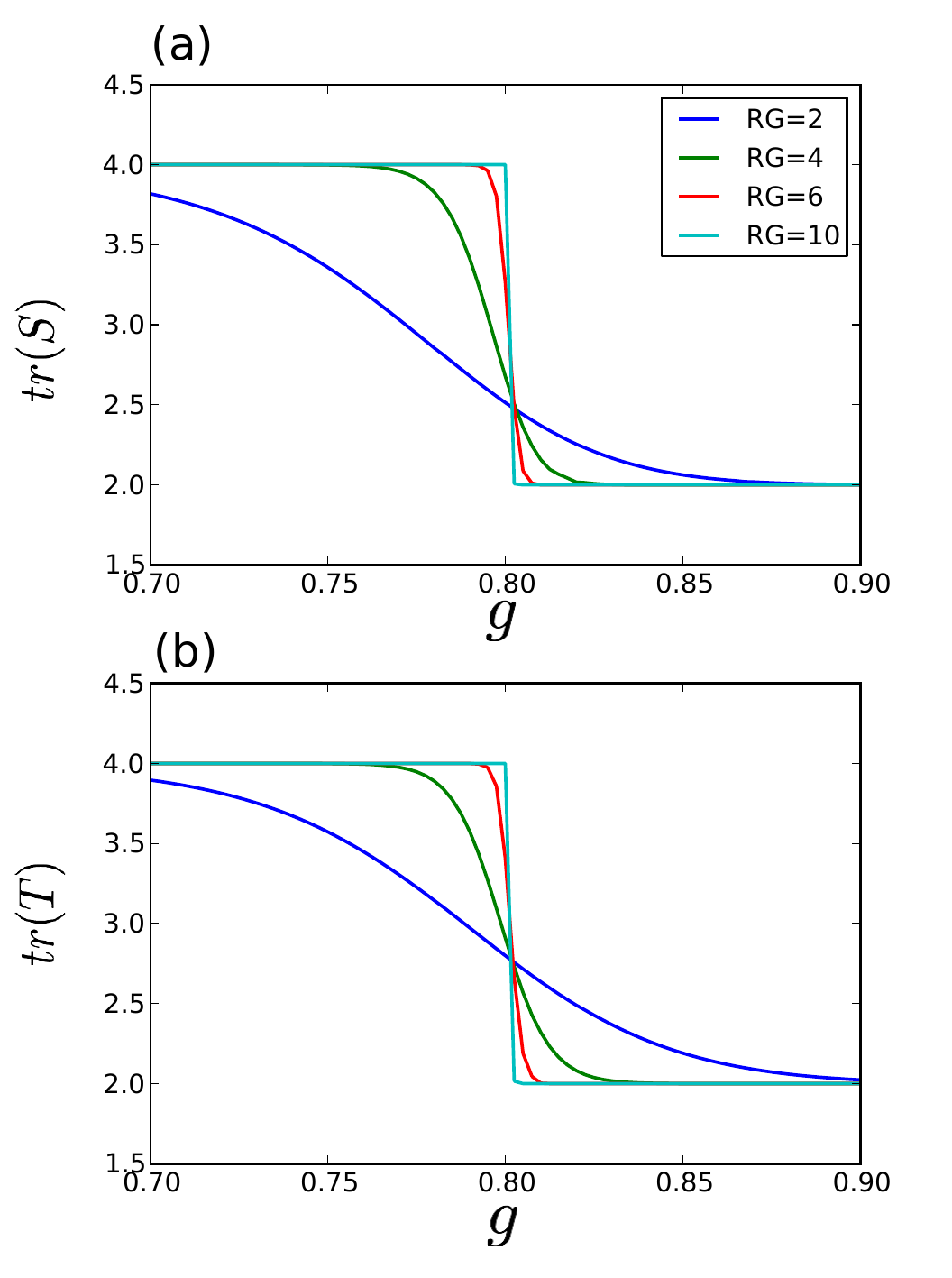,angle=0,width=8cm}}
\caption{  The $\mathbb{Z}_2$ toric code model: the trace of modular matrices (a) $S$ and (b) $T$ as functions of parameter  display a phase transition at critical point $g_c=0.802$ under the renormalization flow.} 
 \label{fig:Z2_topological}
\end{figure}

\smallskip
\noindent {\bf The toric code phase\/}.
Let us first consider the 2D $\mathbb{Z}_2$ toric code wavefunction with deformation $g$, which is represented by the tensor product state $A^{i,j,k,l}_{\alpha,\beta,\gamma,\delta}$ with four physical indices $i,j,k,l=0,1$ and four virtual indices $\alpha,\beta,\gamma,\delta=0,1$. 
The tensor's entries are given by 
\begin{align} A_{i,j,k,l}^{i,j,k,l} = \label{eqn:ToricCode} 
  \begin{cases}
    g^{i+j+k+l}      & \quad \text{if } i+j+k+l=0\, \text{mod }  2,\\
    0  & \quad  \text{otherwise.}\\
  \end{cases}
\end{align}

The parameter $g$,  an effective string tension, is used to tune the property of this state from the topological phase (when $g$ close to $1$) to a trivial phase (when $g$ close to $0$). 
Even though the wavefunction is continuous in $g$, as we vary $g$ the state must undergo a phase transition. In Ref.~\cite{Chen_Gu_Wen_2DQSRG_2010,He_Moradi_wen_GPTRG_2014,Huang_Wei_Z3TO_2015}, it was shown that the phase transition occurs at $g \approx 0.802 $, which separate the $\mathbb{Z}_2$  topologically ordered phase from a trivial phase.

The  MPO corresponds to the above family of wavefunctions~(\ref{eqn:ToricCode}) is generated by product of $Z = \bigl(\begin{smallmatrix}
1&0 \\ 0&-1
\end{smallmatrix} \bigr)$ to internal indices~\cite{He_Moradi_wen_GPTRG_2014}, independent of the parameter $g$.
With this we can now apply the HOTRG to coarse-grain the local tensors and the matrix product operators which can be inserted to TPS to obtain  the degenerate ground states. 
We find that for $0<g \leq 0.802$  the modular matrices are trivial, 
\begin{align}
\label{trivial_oZ2}
S =T=  \begin{pmatrix}
  1 & 1&1 &1 \\
  1& 1&1 &1\\
   1 & 1&1 &1\\
   1 & 1&1 &1
 \end{pmatrix},
\end{align}
 which have only one nonzero eigenvalue 1. (We note that the error is about $10^{-10}$ or smaller.) 
 When  $0.802<g <1$, the state belongs to the $\mathbb{Z}_2$ topologically ordered phase, since we obtain nontrivial $S$ and $T$ matrices as follows,
\begin{align}
\label{nontri_oZ2}
S =  \begin{pmatrix}
  1 & 0&0 &0 \\
  0& 0&1 &0 \\
   0 & 1&0 &0\\
   0 & 0&0 &1
 \end{pmatrix},  \ \
 T =  \begin{pmatrix}
  1 & 0&0 &0 \\
  0& 1&0 &0 \\
   0 & 0&0 &1\\
   0 & 0&1 &0
 \end{pmatrix}.
\end{align}
The results agree as those in Ref.~\cite{He_Moradi_wen_GPTRG_2014}. 
Note that in the topological charge basis, the modular $T$ matrice is diagonal with  each diagonal element of the form $e^{i\theta_s}$ that gives the self-statisitcs of anyonic excitations. We can diagonalize $T$ by a unitary matrix $M$ and at the same time transform $S$ to the same basis~\cite{Zhang_Ashvin_quasiparticle_2012,Liu_STmatrix_2013} 
\begin{align}
\label{nontri_oZ2B}
T' =  \begin{pmatrix}
  1 & 0&0 &0 \\
  0& 1&0 &0 \\
   0 & 0&1 &0\\
   0 & 0&0 &-1
 \end{pmatrix},  \ \  S'  =  \frac{1}{2}\begin{pmatrix}
  1 & 1&1 &1 \\
  1& 1&-1 &-1 \\
   1 & -1&1 &-1\\
   1 & -1&-1 &1
 \end{pmatrix} .
\end{align}

  Our numerical results, using $tr(\hat{S})$ and $tr(\hat{T})$ as the order parameters, are shown in Fig.~\ref{fig:Z2_topological}. There, we also show the values of the order parameters for a few different number of RG steps. 
We see that as we perform more steps of HOTRG, the crossover-like curves become shaper and shaper and they all cross at  a transition with $g_c=0.802$, which  separates the topological  phase $(g>g_c)$ that has the same modular matrices (\ref{nontri_oZ2}) from the non-topological one $(g<g_c)$ with trivial modular matrices (\ref{trivial_oZ2}).

\begin{figure}[t]
\center{\epsfig{figure=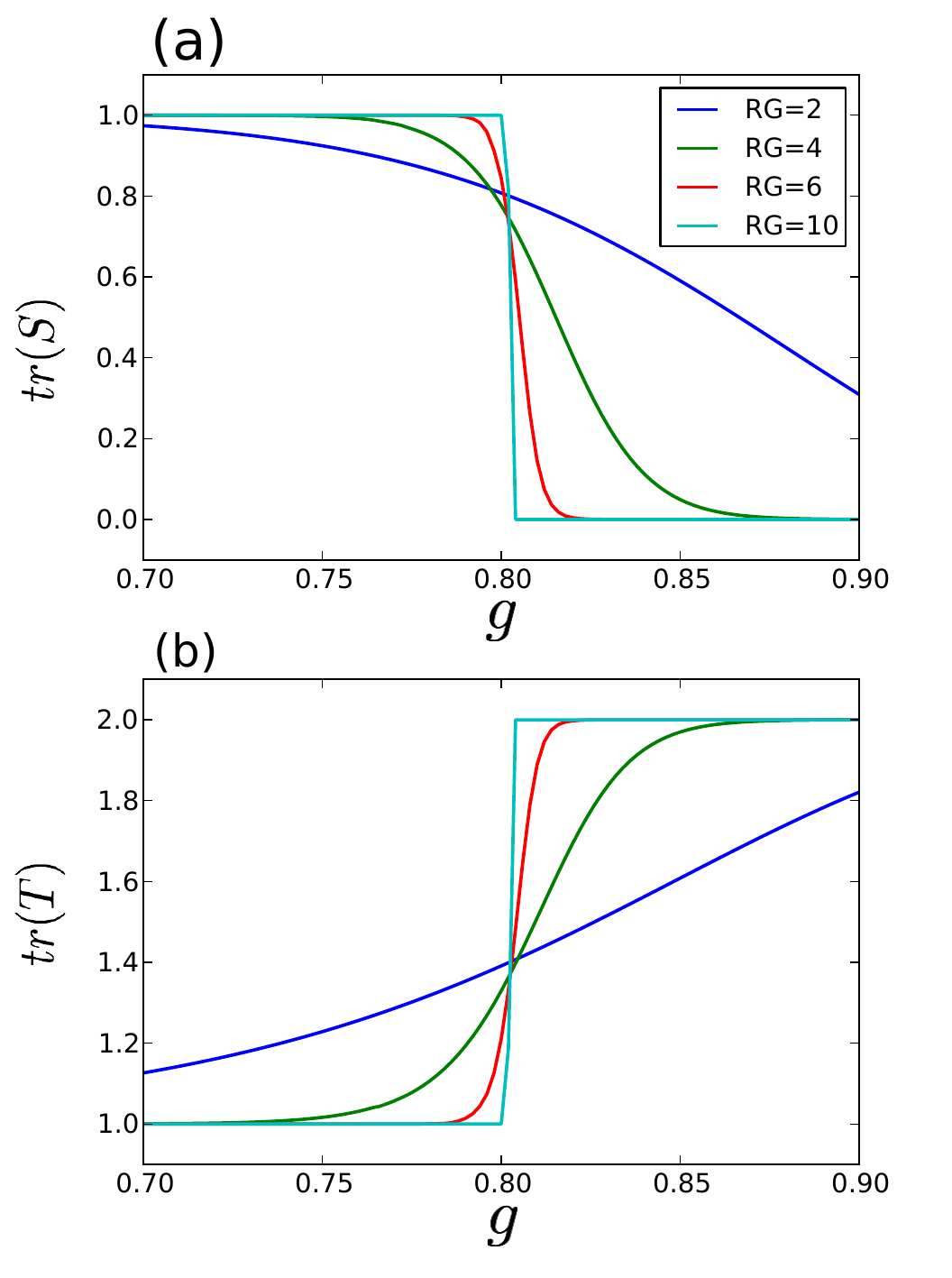,angle=0,width=8cm}}
\caption{ The  double-semion model: the trace of modular matrices (a) $S$ and (b) $T$ as functions of parameter  display a phase transition at critical point $g_c =0.802$ under the renormalization flow.} 
 \label{fig:double_semion}
\end{figure}

\medskip
\noindent {\bf The double-semion phase\/}.
The double-semion model is another $\mathbb{Z}_2$ topologically ordered phase (i.e. the twisted version of the toric code) with two semions of statistical spin $\theta_s= \pm \pi/2$.  
The wavefunction of the double-semion ground state is a superposition of closed-loop configurations with different weight, $|\Psi\rangle = \sum_{c}(-1)^{N_{loop}}  |\phi_{c} \rangle  $, where $|\phi_{c} \rangle $ represents  closed loops and the $N_{loop}$ the total number of loops for a given configuration $\phi_c$. 
This state and its deformation (parameterized by $g$) can described by a tensor product state~\cite{TNstringnet,TNstringnet2} that is characterized by the rank-eight tensor $ P_{\alpha\alpha'  \beta \beta'  \gamma \gamma' \delta \delta'}$, with internal indices running over $\{0,1\}$ on the vertex
and the rank-five tensor $G^s_{\alpha\alpha'\beta\beta}$ with one physical index $s$ running over the possible spin states $\{0,1\}$~\cite{He_Moradi_wen_GPTRG_2014}:
\begin{align}
|\psi(g) \rangle = \sum_{s_i} tTr (\otimes_v P \otimes_l G^{s_i})|s_1,s_2,....\rangle,
\end{align}
where $v$ labels vertices and $l$ links. Moreover, the projection $P$ is given as 
\begin{align}
 &P_{\alpha\alpha'  \beta \beta'  \gamma \gamma' \delta \delta'}  =  p_ {\alpha  \beta  \gamma \delta}\, \delta_{\alpha \alpha'}  \delta_{ \beta\beta'}   \delta_{\gamma \gamma'}  \delta_{\delta \delta'}  \notag \\
 &p_{1000} = p_{1101} = -1, \quad \text{otherwise}  \quad  p_ {\alpha  \beta  \gamma \delta}=1,
 \end{align}
and  the $G$ tensor is explicitly written as
\begin{align}
&  G^s_{\alpha\alpha' \beta\beta'}= g^s_{\alpha\beta}  \delta_{\alpha \alpha'}  \delta_{ \beta\beta'}  \\
& g^1_{10}=g^1_{01} = g,  \quad g^0_{00}=g^0_{11} = 1, \ \text{and}  \ g^s_{\alpha\beta} =0 \ \text{otherwise}.\notag
 \end{align}
 At $g=1$, this state is the exact fixed-point wavefunction of the double-semion model.  But  at $g=0$, the state is a product state of all $0$'s, i.e. $|0\dots0\rangle$. Thus
by tuning the string tension $g$, a phase transition must occur.  To detect such a transition, we shall characterize phases using modular matrices, and we expect that within the same phase their values will remain the same. The transition can be detected at the location where the values of modular matrices make a change.

In order to determine the modular matrices, we first identify the gauge symmetry transformation for the tensor product state to find the degenerate ground states. 
The  $\mathbb{Z}_2$ gauge transformation is generated by $X \otimes X \alpha$ acting on the each internal indices~\cite{He_Moradi_wen_GPTRG_2014}, where 
$X = \bigl(\begin{smallmatrix}
0&1 \\ 1&0
\end{smallmatrix} \bigr)$, 
and
\begin{align}
\alpha = | 00\rangle \langle 00 | +  i | 01\rangle \langle 01 | + i | 10\rangle \langle 10 | + | 11\rangle \langle 11 |.
\end{align}

With the MPO, we can carry out the evaluation of the modular matrices using HOTRG, and 
we find that there is a phase transition at $g_c=0.802$, as shown in Fig.~\ref{fig:double_semion}.
The $S$ and $T$ matrices for the nontrivial double-semion phase are listed below, 
\begin{align}
\label{nontri_Z2}
S =  \begin{pmatrix}
  1 & 0&0 &0 \\
  0& 0&1 &0 \\
   0 & 1&0 &0\\
   0 & 0&0 &-1
 \end{pmatrix},  \ \
 T =  \begin{pmatrix}
  1 & 0&0 &0 \\
  0& 1&0 &0 \\
   0 & 0&0 &-1\\
   0 & 0&1 &0
 \end{pmatrix},
\end{align}
with $tr(S)=0$ and $tr(T)=2$. In the basis where $T$ is diagonal (and the first column and first row of $S$ being $1/2$), they are
\begin{align}
 T' =  \begin{pmatrix}
  1 & 0&0 &0 \\
  0& 1&0 &0 \\
   0 & 0&i &0\\
   0 & 0&0 &-i
 \end{pmatrix},  \ \ S' =  \frac{1}{2}\begin{pmatrix}
  1 & 1&1&1 \\
  1& 1&-1 &-1 \\
   1 & -1&-1 &1\\
   1 & -1&1 &-1
 \end{pmatrix} .
\end{align}
In contrast, for the trivial phase, the modular matrices are
\begin{align}
\label{trivial_Z2}
S=T =  \begin{pmatrix}
  1 & 0&0 &0 \\
  0& 0&0 &0 \\
   0 & 0&0 &0\\
   0 & 0&0 &0
 \end{pmatrix}.
\end{align}
Thus, the modular matrices are capable of characterizing phases and detecting transitions.

We note that the trivial modular matrices here actually differ from those in Eq.~(\ref{trivial_oZ2}) for the deformed toric code.  Why is there such a difference? 
The gauge transformation of the double-semion phase is actually $X$, and it can be used to create a domain wall along the twist line. 
Since there is no degenerate ground state for the trivial phase,  once a domain wall is created, the tensors  of twisted wave function cannot be `contracted' to obtain nonzero values.  
Therefore,  there is  only one nonzero term $ \big{\langle} \Psi( \mathbb{I},\mathbb{I}  ) |  \Psi(  \mathbb{I},\mathbb{I} )  \big{\rangle}$ in the modular matrices.
In contrast, the gauge transformation of the deformed toric code is $Z$, and it does not create a domain wall but only a phase factor. Any combination of the symmetry twists give identical states and thus we obtain a different form of trivial modular matrices.

We would like to emphasize that, in carrying out the above procedure, we did not impose the gauge symmetry, as was done in Ref.~\cite{He_Moradi_wen_GPTRG_2014}. Imposing such symmetry can in principle make the numerical procedure stable, but our HOTRG-based tnST approach demonstrated here already yields very stable values along the RG flow. The fact that we do not need to impose additional symmetry bodes well for the SPT phases to be described below, as in that case there is  no such a gauge symmetry to be exploited and only the symmetry associated with action on physical degrees of freedom.

\begin{figure}[t]
\center{\epsfig{figure=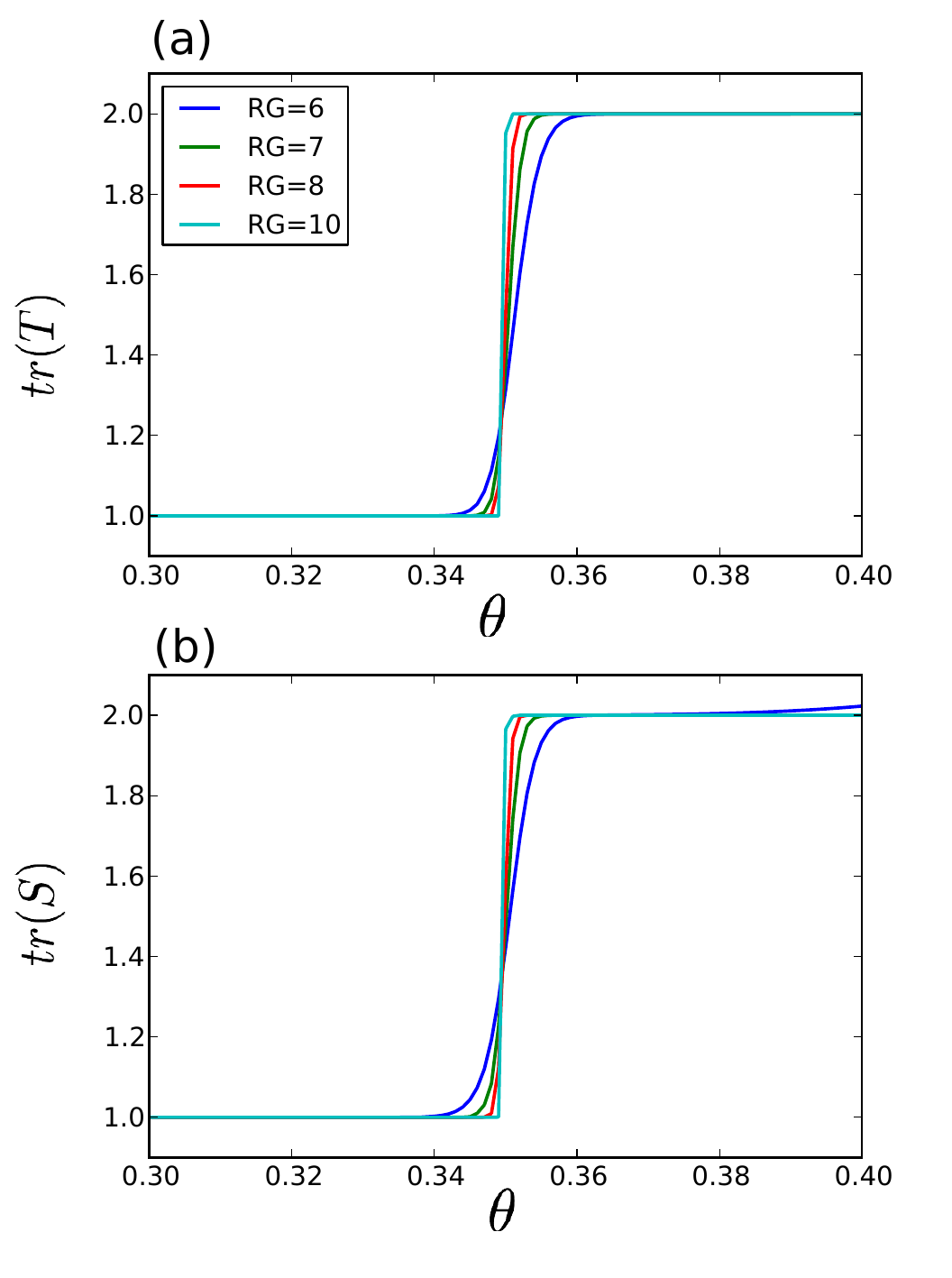,angle=0,width=8cm}}
\caption{ The Ising PEPS model: the trace of modular matrices (a) $S$ and  (b) $T$ as functions of parameter  display a phase transition at critical point $\theta_c=0.349$ under the renormalization flow.} 
 \label{fig:ising_peps}
\end{figure}

\subsection{ Long-range order and spontaneous symmetry breaking: the Ising PEPS}
\label{subsec:SB}
Having seen that our tnST method works for identifying topological phases, here we demonstrate that  it also works for detecting a transition between symmetry-breaking and symmetric phases, corresponding to classical ordered and disordered phases, respectively. 
In particular, the model we are about to consider is the Ising PEPS \cite{Verstraete_PEPS_2006,Norbert_SB_PEPS_2015} on the square lattice  with tensor $A=|0\rangle \langle \theta , \theta , \theta,  \theta  |+ | 1\rangle \langle \bar{\theta} , \bar{\theta} , \bar{\theta},  \bar{\theta} |$, where the ket (bra) corresponds to the physical (virtual) degrees of freedom, and $ |\theta \rangle= \cos \theta  |0\rangle  +\sin \theta  |1\rangle$ and $ |\bar{\theta} \rangle= \sin \theta  |0\rangle  +\cos \theta  |1\rangle$ with $\theta\in [0,\pi/4]$.  A corresponding local Hamiltonian can be written down (not shown here)~\cite{Verstraete_PEPS_2006}.
This model has  a  $\mathbb{Z}_2$  symmetry with the nontrivial group action being the Pauli matrix $\sigma_x$ on the physical qubit. 
The above tensor represents the following state with coefficients related to the Boltzmann weight  (more precisely, its square root) of the classical model,
\begin{align}
|\Psi \rangle = \sum_{\{s_i=\pm 1\}}e^{-\beta/2 H_{\rm Ising}(s_1,s_2,...,s_n) } |q_1,q_2,...,q_n\rangle,
\end{align}
where $H_{\rm Ising}(s_1,s_2,...,s_n) = -\sum_{\langle i,j\rangle} s_i s_j$ is the 2D classical Ising model with the classical variables $s_i=\pm 1$ and the corresponding quantum bits $q_i$'s in the Pauli Z basis being $q_i=(1-s_i)/2=0/1$. 
In Ref.~\cite{Verstraete_PEPS_2006}, it was shown that, by the correspondence of the quantum ground state to the classical Ising model  via $\sin2\theta= e^{-\beta}$, there is a second-order quantum phase transition occurring at $\theta_c \approx 0.349596$. This model was also recently studied with TN methods for investigating how symmetry breaking emerges even though the PEPS description is symmetric~\cite{Norbert_SB_PEPS_2015}. In this regard, the modular matrices to be presented below provide an alternative approach.

\begin{figure}[t]
\center{\epsfig{figure=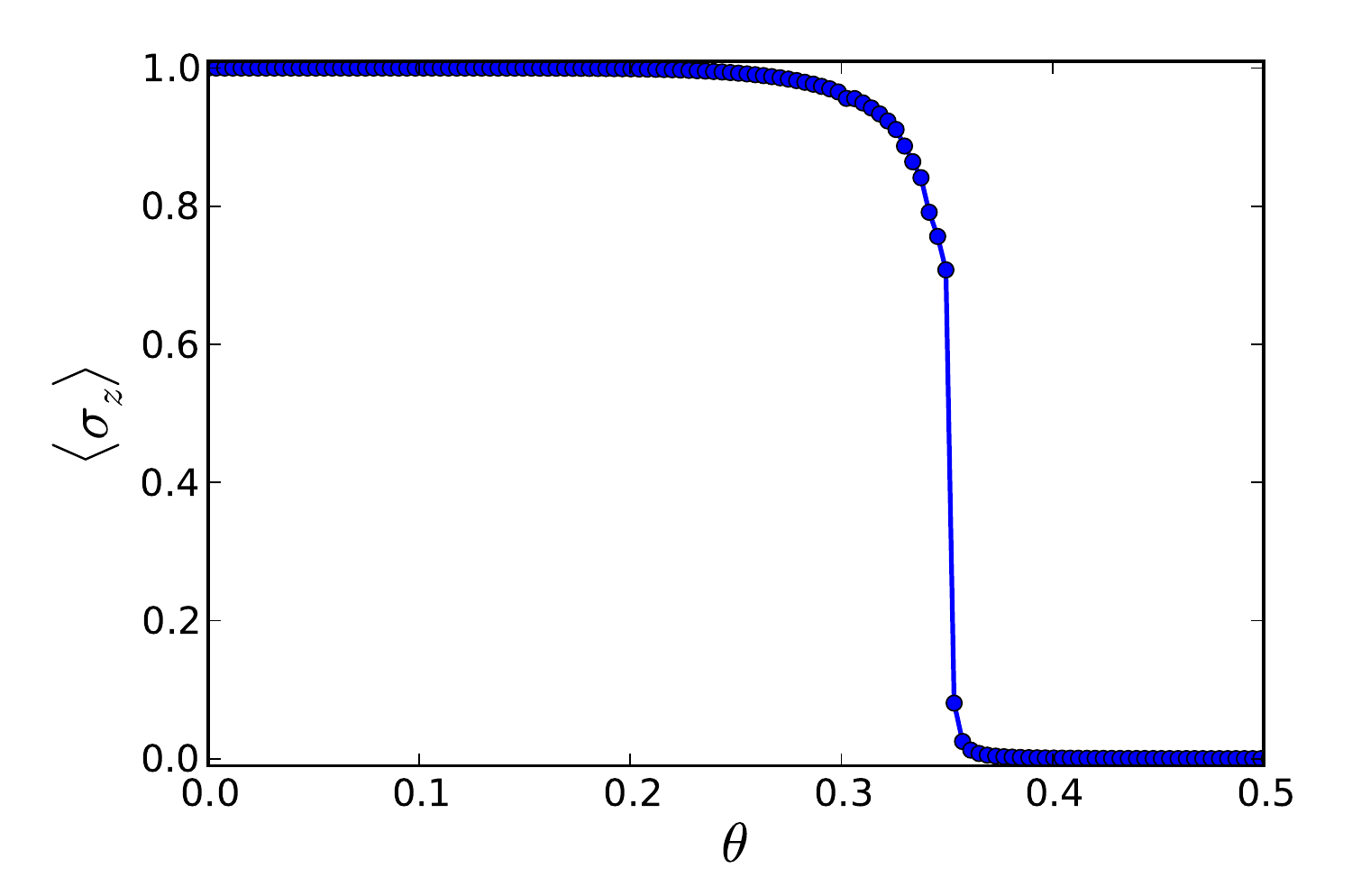,angle=0,width=7cm}}
\caption{ The Ising PEPS model: the magnetization as functions of parameter  display a phase transition at critical point $\theta_c=0.349$ with RG step is 10 and $D_{cut}=16$.} 
 \label{fig:ising_peps_Sz}
\end{figure}

Here we apply our method of modular matrices, and their calculation can be performed similarly as in the topological models, except that we need to use the symmetry twists generated by the matrix product operator $X \otimes X \otimes ... \otimes X$ 
with 
$X = \bigl(\begin{smallmatrix}
0&1 \\ 1&0
\end{smallmatrix} \bigr)$
acting on  the internal indices.
We  obtain the modular matrices by varying $\theta$ from $0$ to $0.5$ as shown in  Fig.~\ref{fig:ising_peps}. 
At $0\le\theta\leq 0.349$, the state has  trivial modular matrices [see Eq.~(\ref{trivial_Z2})]
and it is actually a spontaneous symmetry-breaking phase and we can understand this by examining, for example, the tensor at $\theta=0$, which gives the quantum state $|\psi \rangle = | 00....0\rangle +|11...1\rangle$. Even though it is of the Schr\"odinger-cat form, small physical fluctuation will break the $\mathbb{Z}_2$ symmetry and the ground state will select spontaneously either $|00...0\rangle$ or $|11...1\rangle$, displaying long-range order. To verify the long-range order we also use HOTRG to calculate $\langle \sigma_z\rangle$ averaged over spins and present the results in Fig.~\ref{fig:ising_peps_Sz}.
The magnetization starts with unity at $\theta=0$ (corresponding to $T=1/\beta=0$) and sharply drops to zero at $\theta_c \approx 0.349$ (corresponding to $T\approx 2.269$), agreeing with the result from the modular matrices.
On the other hand when $0.349<\theta\leq \pi/4$ (with the plot cut off at $\theta=0.5$ for convenience),  we obtain the non-trivial modular matrices [see also Eq.~(\ref{nontri_oZ2})], and this is a symmetric phase (albeit a trivial SPT phase) that corresponds to the classical disordered phase. 

We remark that as $\theta$ approaches $\theta_c$, the curves for $tr(S)$ and $Tr(T)$ in Fig.~\ref{fig:ising_peps} show a crossing, and from this we obtain the critical exponent $\nu \approx 1.0$~\cite{Huang_Wei_Z3TO_2015}; see also Fig.~\ref{fig:Z2_TRI_CP} and Fig.~\ref{fig:Z3_TRI_CP} below for similar analysis.  On the other hand,  we also calculate directly the correlation function right at the critical point: $\langle \sigma_z(R_i) \sigma_z(R_j)  \rangle \propto |R_i-R_j|^{-d+2-\eta}$ and obtain $\eta \approx 0.23$, which is close to $\eta=1/4$ for the 2D Ising model.

From this and previous examples, we have seen that the modular matrices provide  good order parameters to distinguish  symmetry-breaking, symmetric and topologically ordered phases. But can they also be used to detect nontrivial symmetric phases?

\subsection{2D symmetry breaking phase to  $\mathbb{Z}_N$ SPT phase}\label{subsec:ZNSPT} 

\begin{figure}[t]
\center{\epsfig{figure=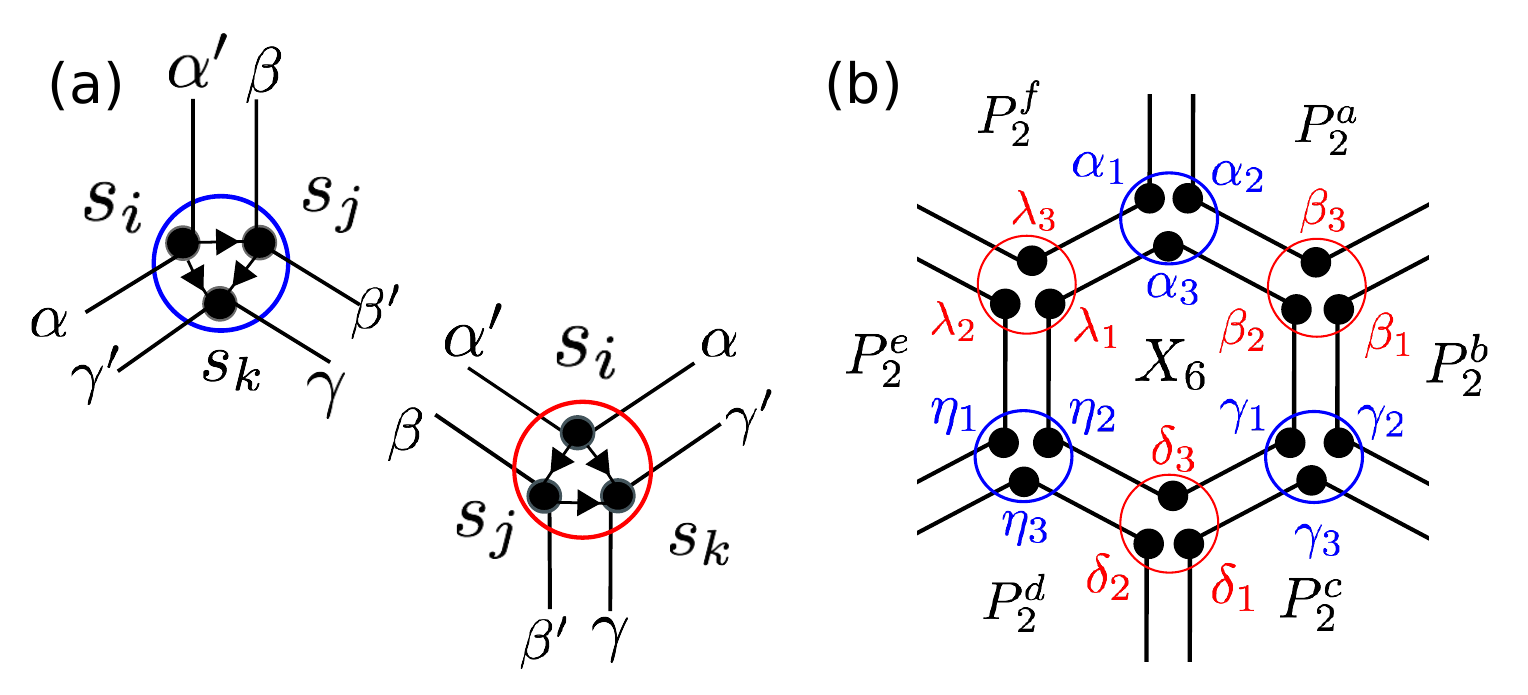,angle=0,width=8cm}}
\caption{ (a) The graphic representation of the  $\mathbb{Z}_N$ SPT and the phase factor $\nu_3^{\pm 1}(s_i,s_j,s_k,g^*)$ in Eq.~\ref{eq:ZNTPS}. Each site contain three particle.
(b) A local term in the Hamiltonian, which is tensor product of one six-site operator $X_6$ and six two-site operators $P_2$ on honeycomb lattice.  }
  \label{fig:ZN_tensor}
\end{figure}

Let us begin by describing the construction of  2D  SPT  wavefunctions through group cocycles of a group $G$~\cite{Chen_cohomology_2013}. 
Consider the physical system  on the honeycomb lattice, with each singe site containing three particles or partons, as shown in Fig~\ref{fig:ZN_tensor}(a).  
The tensor product state for the fixed-point SPT state is  characterized by a  tensor $A^{(s_i,s_j,s_k)}_{(\alpha,\alpha',\beta,\beta',\gamma,\gamma')}$ with three physical indices and six internal indices all running over all group elements $g\in G$,  
\begin{align}
|\Psi \rangle = \sum_{s_i} tTr (A \otimes A... \otimes A) |s_1,s_2,...\rangle,
\end{align}
where the tensor  $A^{(s_i,s_j,s_k)}_{(\alpha,\alpha',\beta,\beta',\gamma,\gamma')}$ is related to the 3-cocycle of $G$  and whose indices satisfy that $s_i=\alpha=\alpha'$, $s_j= \beta=\beta'$, and  $s_k=\gamma=\gamma'$:
\begin{eqnarray}
\label{eq:ZNTPS}
&&A^{(s_i,s_j,s_k)}_{(\alpha,\alpha',\beta,\beta',\gamma,\gamma')}\\
&=&\nu_3^{\pm 1}(s_i,s_j,s_k,g^*)\delta_{s_i,\alpha}\delta_{s_i,\alpha'}\delta_{s_j,\beta}\delta_{s_j,\beta'}\delta_{s_k,\gamma}\delta_{s_k,\gamma'},\nonumber
 \end{eqnarray} 
where $\nu_3(s_i,s_j,s_k,g^*)$ is the 3-cocycle of the group $G$, the exponent $\pm1$ depends whether the triangle formed by the three partons is upside down ($+1$) or not ($-1$), and $g^*$ is a fixed element in $G$. Since the virtual indices are identical to the associated physical indices, we can simply denote the tensor as $A{(s_i,s_j,s_k)}$. In the following we shall mainly consider the symmetry group $G=\mathbb{Z}_N$, whose symmetry action is 
generated  by the operator $\hat{X} = \sum_{i=0}^{N-1} | \text{mod}(i+1,N) \rangle \langle i |$ acting on all partons on every site. For $\mathbb{Z}_N$ 
there are $N$ inequivalent group cohomology classes in $H^3(\mathbb{Z}_N,U(1))$, labeled by
 $k=0,1,2,...,N-1$. 
 From these fixed-point wavefunctions we shall construct a family of them such that the different SPT phases labeled by different $k$'s are continuously deformed into one another and investigate phase transitions that should arise. 

\subsubsection{A $Z_2$ example: SPT phases separated by a symmetry-breaking phase} 
\begin{figure}[ht]
\center{\epsfig{figure=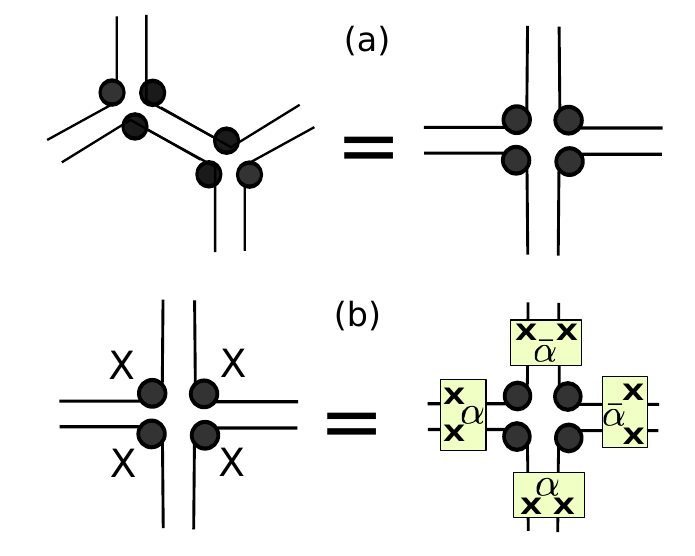,angle=0,width=6cm}}
\caption{ (a) Combine every two sites and map the system on the square lattice. (b) Local  $\mathbb{Z}_N$ symmetry action acts on each site.  }
  \label{fig:ZN_MPO}
\end{figure}

Here we consider $\mathbb{Z}_2$ symmetry. The tensor for the fixed-point wavefunction $|\Psi^{(k)}\rangle$ is  given by $A^{(s_i,s_j,s_k)}_{(\alpha,\alpha',\beta,\beta',\gamma,\gamma')} \equiv A(s_i,s_j,s_k)$ with 
\begin{align}
\label{Z2SPT_TPS}
& A(0,0,0) = A(1,1,1) = 1 \notag \\
& A(0,0,1) = A(0,1,0) = A(1,0,0) = 1 \notag \\
& A(1,1,0) = A(1,0,1) = A(0,1,1) = (-1) ^k,
\end{align}
where $k=0$ labels the trivial  $\mathbb{Z}_2$ SPT phase and $k=1$ the nontrivial one. 
It is easy to verify that these wavefunctions are invariant under the $\mathbb{Z}_2$ action generated by
 the operator ${X} =  |0\rangle\langle 1|+|1\rangle\langle 0|$ on all partons.
However, the transformed tensors differ from the original ones by a local unitary transformation or MPO on the inner indices which is given by $X \otimes X \alpha$, as shown in Fig.~\ref{fig:ZN_MPO}(b), where 
\begin{align}
\alpha = \bar{\alpha}= | 00\rangle \langle 00 | +  | 01\rangle \langle 01 | + (-1)^k| 10\rangle \langle 10 | + | 11\rangle \langle 11 |.
\end{align}
This also gives a way of verifying that the wavefunction is $\mathbb{Z}_2$ symmetric, as the MPO's from neighboring sites cancel one another.

\medskip
We now construct one-parameter family of $\mathbb{Z}_2$ symmetric wavefunctions that connect the two fixed-point wavefunctions $|\Psi^{(0)}\rangle$ and $|\Psi^{(1)}\rangle$; see Eq.~(\ref{Z2SPT_TPS}). To do this we consider a deformation $Q(g)$ acting on all three partons on each site, 
\begin{eqnarray}
\label{eqn:DeformationQ}
Q(g)&=&\sum q_{s_is_js_k} |s_is_js_k \rangle \langle s_is_js_k|\\
&=&|000\rangle\langle 000| + |111\rangle\langle 111|\nonumber \\
&& + |g|\sum_{(s_i,s_j,s_k)\ne 000,111} |s_i,s_j,s_k\rangle\langle s_i,s_j,s_k|\nonumber,
\end{eqnarray}
so that 
\begin{equation}
|\Psi^{(k)}(g) \rangle \equiv Q(g)\otimes Q(g) \otimes... \otimes Q(g) |\Psi^{(k)} \rangle. 
\end{equation}
Such a deformation manifestly preserves the $\mathbb{Z}_2$ symmetry. But we observe that at $g=0$, $|\Psi^{(0)}(0)\rangle = |\Psi^{(1)}(0)\rangle = |000\dots000\rangle + |111\dots111\rangle$ is a cat state.
Therefore, the two branches of the deformed states are smoothly connected at $g=0$ and we can define the one-parameter family of states to be
\begin{equation}
\label{eqn:Z2WF}
|\Psi(g)\rangle\equiv\begin{cases}
    |\Psi^{(1)}(g)\rangle     & \quad \text{if } g<0,\\
    |\Psi^{(0)}(g)\rangle  & \quad  \text{if } g\ge 0.\\
  \end{cases}
\end{equation}
The corresponding tensors are thus given by
\begin{align}
\label{Z2_SPT_deform}
& A(0,0,0) = A(1,1,1) = 1 \notag \\
& A(0,0,1) = A(0,1,0) = A(1,0,0) = |g| \notag \\
& A(1,1,0) = A(1,0,1) = A(0,1,1) = g.
\end{align}

\begin{figure}[t]
\center{\epsfig{figure=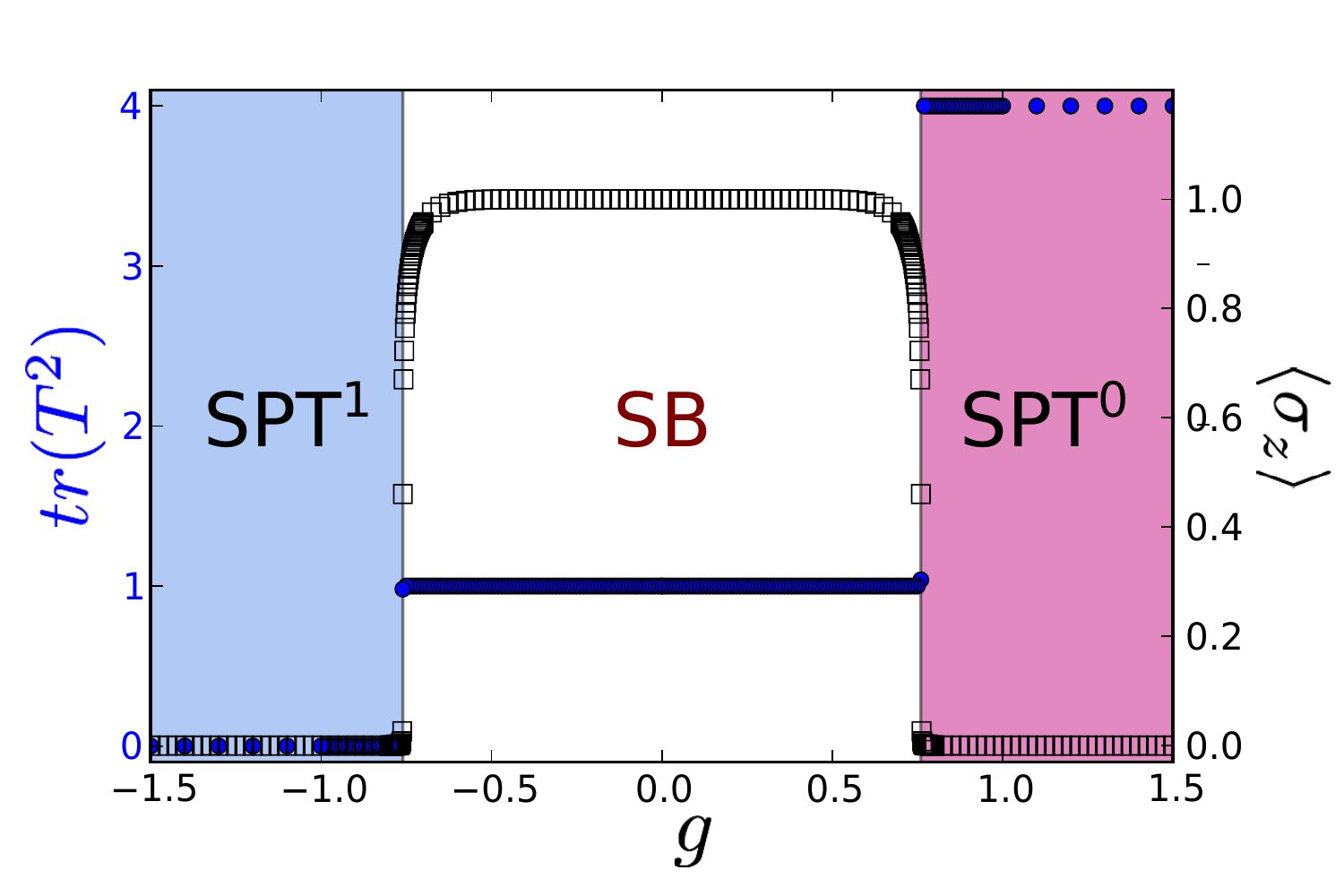,angle=0,width=8cm}}
\caption{  The $\mathbb{Z}_2$  SPT model represented by tensor [Eq.~(\ref{Z2_SPT_deform})]: 
the trace of  $T^2$ (solid blue circle)  and the magnetization $\langle \sigma_z \rangle$ (black empty square) as functions of parameter $g$  display the phase transitions at critical point $g_{c1}=-0.760$ from non-trivial  $\mathbb{Z}_2$  SPT labeled as $\text{SPT}^1$  to symmetry breaking phases labeled as SB and at critical point $g_{c2}=0.760$ from symmetry breaking phase to the trivial $\mathbb{Z}_2$  SPT labeled as $\text{SPT}^0$ with $D_{cut}=16$ and the step of HOTRG is 10. }
  \label{fig:Z2_TRI}
\end{figure}

We thus see that by tuning the parameter $g$ continuously,  two different $\mathbb{Z}_2$ SPT phases can be  connected and there is a symmetry-breaking cat state in between (at $g=0$). But is the transition point located at $g=0$ or somewhere else? What is the phase diagram? 

We remark that the wavefunction~(\ref{eqn:Z2WF}) is the ground state of the following Hamiltonian of Eq.~(\ref{app:Z2WF}) with energy identically zero,
\begin{equation}
%\label{app:Z2WF}
H(g) \equiv\begin{cases}
    \sum_p    \mathbb{R}_p(g)   \  h_1^{\{p\}} \  \mathbb{R}_p(g)    & \quad \text{if } g<0,\\
    \sum_p    \mathbb{R}_p(g)  \ h_0^{\{p\}}  \ \mathbb{R}_p(g)     & \quad  \text{if } g\ge 0,\\
  \end{cases}
\end{equation}
where the detail of its construction is explained in Appendix~\ref{sec:firstZ2}. We expect that any transition arising from this will be continuous.

We would like to emphasize that even though the MPO is the same as that for the fixed-point one in the same branch of the wavefunction,  in our numerical calculations we check which MPO makes the state invariant and then use that MPO for evaluating the modular matrices.  
Moreover, for simplicity of calculation, we merge every two sites on the hexagonal lattice  and map the system to the square lattice  as shown in Fig.~\ref{fig:ZN_MPO}(a)~\cite{Hendrik}. Then each local tensor has four double  inner indices and four physical indices.

The resulting phase diagram is shown in Fig.~\ref{fig:Z2_TRI} and it turns out that there are two phase transitions, with a finite region of symmetry-breaking phase separating the two SPT phases. 
When $-0.760 \leq g < 0.760$, we obtain trivial modular matrices
\begin{align}
\label{T2matrix_tri}
S = T= \begin{pmatrix}
  1 & 0&0 &0 \\
  0& 0&0 &0 \\
   0 & 0&0 &0\\
   0 & 0&0 &0
 \end{pmatrix},
\end{align}
and this shows that this region is the symmetry-breaking phase, same as the cat state. 
When $g>0.760$,  we obtain the following modular matrices
\begin{align}
\label{nontri_Z2}
S =  \begin{pmatrix}
  1 & 0&0 &0 \\
  0& 0&1 &0 \\
   0 & 1&0 &0\\
   0 & 0&0 &1
 \end{pmatrix},  \ \
 T =  \begin{pmatrix}
  1 & 0&0 &0 \\
  0& 1&0 &0 \\
   0 & 0&0 &1\\
   0 & 0&1 &0
 \end{pmatrix}.
\end{align}
The main characterization of the SPT phase ($\mathbb{Z}_N$ in general) is  
the $T^2$ matrix,
\begin{align}
 \label{T2matrixk0}
T^2 =  \begin{pmatrix}
  1 & 0&0 &0 \\
  0& 1&0 &0 \\
   0 & 0&1 &0\\
   0 & 0&0&1 
 \end{pmatrix},
\end{align}
whose trace is $4$.   Thus this region belongs to the trivial $\mathbb{Z}_2$ ($\text{SPT}^0$).
For  $g<-0.760$, we obtain the  following modular matrices,
\begin{align}
\label{nontri_Z2}
S =  \begin{pmatrix}
  1 & 0&0 &0 \\
  0& 0&1 &0 \\
   0 & 1&0 &0\\
   0 & 0&0 &-1
 \end{pmatrix},  \ \
 T =  \begin{pmatrix}
  1 & 0&0 &0 \\
  0& 1&0 &0 \\
   0 & 0&0 &-1\\
   0 & 0&1 &0
 \end{pmatrix},
\end{align}
the $T^2$ matrix is
\begin{align}
 \label{T2matrixk1}
T^2 =  \begin{pmatrix}
  1 & 0&0 &0 \\
  0& 1&0 &0 \\
   0 & 0&-1 &0\\
   0 & 0&0&-1 
 \end{pmatrix},
\end{align}
whose trace is $0$.  Thus this regions  belongs to the non-trivial $\mathbb{Z}_2$ ($\text{SPT}^1$) symmetry-protected topologically ordered phase. We note that the modular matrices for both SPT phases agree with our analytic results in Sec.~\ref{sec:ST} for the fixed-point SPT wavefunctions. 

\begin{figure}
\center{\epsfig{figure=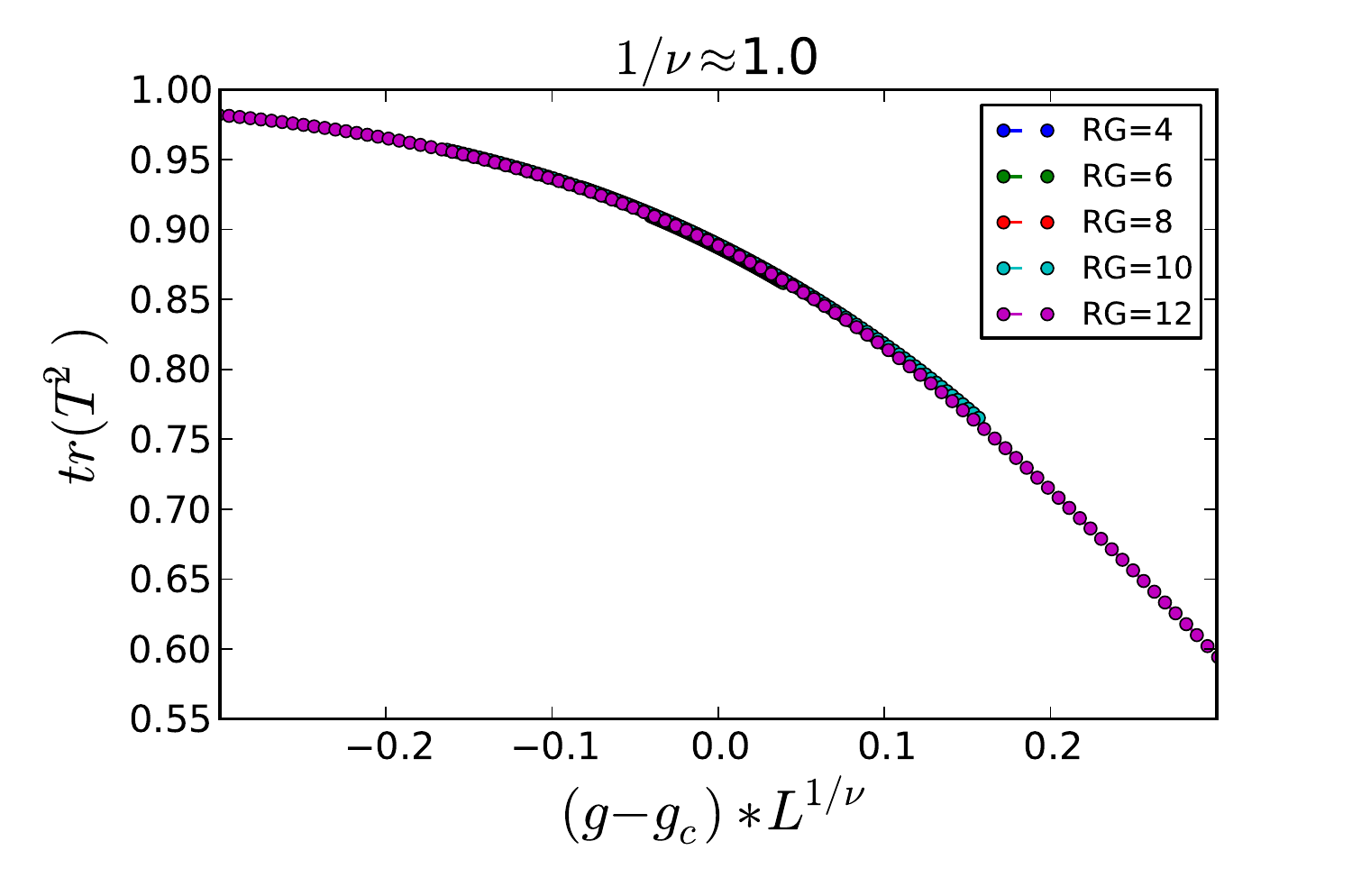,angle=0,width=8cm}}
\caption{ The trace of $T^2$ as a function $(g-g_c)\times L^{\nu}$ on the $\mathbb{Z}_2$  SPT model represented by tensor [Eq.~(\ref{Z2_SPT_deform})] near transition points $g_{c2} \approx 0.760$. }
  \label{fig:Z2_TRI_CP}
\end{figure}
As seen  from Fig.~\ref{fig:Z2_TRI}, the topological order parameters clearly show a sharp change around $|g_c| = 0.760$, and the transition point separates the symmetry-breaking phase from the symmetry protected topological  phase. Also shown in the figure is the spontaneous magnetization $\langle \sigma_z\rangle$ averaged over all partons and it agrees with the characterization of the symmetry-breaking phase and the transitions into the two SPT phases.

We would like to comment that the norm square of the one-parameter state (\ref{eqn:Z2WF}) is proportional to  the partition function of the classical ferromagnetic Ising model on the triangular lattice with the inverse temperature given by $\beta = -{\ln g}/{2}$.
From the transition point of this classical model we infer that $g_c = 1/\sqrt{3} \approx 0.759836$ and it is  a continuous quantum phase transition. 
As the critical point is approached, the correlation length diverges as $\xi  \sim |T-T_c| ^{-\nu}$ for the classical model,  where $\nu$ is the correlation-length critical exponent. 
For  the ferromagnetic Ising model on triangular lattice, the critical exponent $\nu=1$~\cite{Fisher_CPTRI_1967,Mon_CPTRI_1993}, and since our wavefunction norm square maps to the partition of this model  we expects the ground state of our quantum model at criticality has the same exponent in its correlation length, i.e., $\xi\sim |g-g_c|^{-1}$. Indeed, we extract the critical exponent $\nu$ from the data collapse of the order parameter, i.e. $tr(T^2)$, under the renormalization flow~\cite{Huang_Wei_Z3TO_2015}. Since after each step of the renormalization the number of sites is reduced by four, the effective linear size $L$ of the  system doubles after each RG step, giving $L=2^{n_{rg}}$ for $n_{rg}$ steps. By finding the best $\nu$ that gives the data collapse in the $tr(T^2)$ vs. $L^{1/\nu} (g-g_c)$, as shown  
 in Fig.~\ref{fig:Z2_TRI_CP},  we find $1/\nu \approx 1.0 $. 

To sum up, the HOTRG approach accurately determines the phase transition between the symmetry-protected topological order phase and the symmetry-beeaking phase, the result of which matches very well with the mapping to the classical Ising model on the triangular lattice.

\begin{figure}[t]
\center{\epsfig{figure=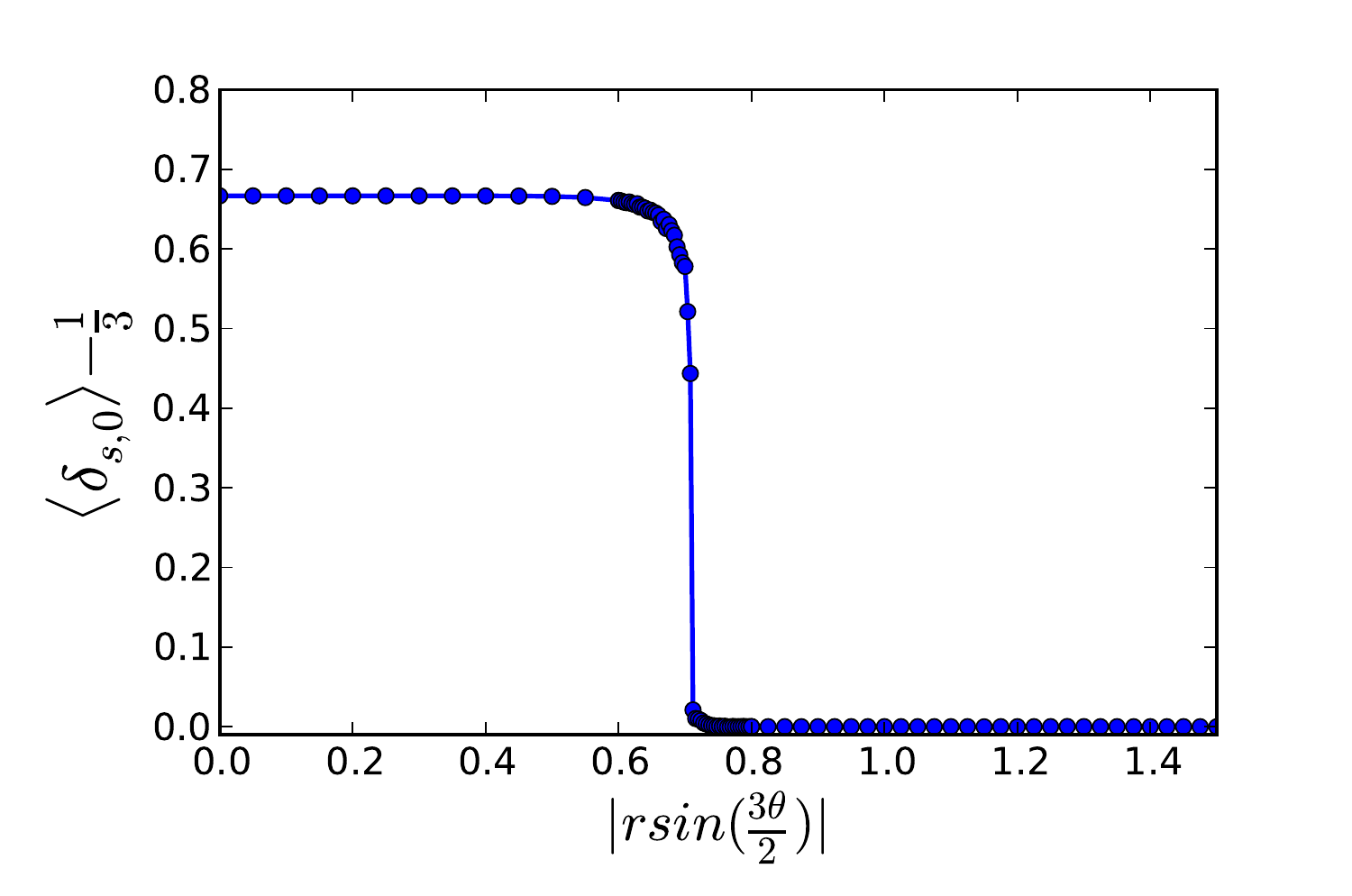,angle=0,width=7cm}}
\caption{ The $\mathbb{Z}_3$  SPT model at $\theta=\pi/3$: the order parameter $\langle \delta_{s,0} \rangle-\frac{1}{3}$ as functions of parameter  display a phase transition at critical point $r_c=0.708$  with $D_{cut}=20$ and 10 RG steps.   } 
 \label{fig:Z3_TRI_MZ}
\end{figure}
\subsubsection{A $Z_3$ example: SPT phases separated by a symmetry-breaking phase}

For  $\mathbb{Z}_3$ SPT phases, the fixed-point wavefunction from the 3-cocycle is given  by the following local tensor 
\begin{align}
&A(0,0,0) = A(1,1,1) =A(2,2,2)= 1 \notag \\
&A(0,1,2) = A(0,2,1) =A(1,1,2) =A(2,2,0) =1 \notag \\
&A(1,2,1) = A(2,0,2) =A(2,1,1) =A(0,2,2) =1 \notag \\
&A(1,2,2) = A(2,0,0) =A(2,1,2) =A(0,2,2) =1 \notag \\
&A(2,2,1) = A(0,0,2) =1\notag \\
&A(2,0,1) = A(2,1,0) =A(0,0,1) = A(1,0,0) =\omega ^k \notag \\
&A(0,1,0) = A(0,1,1) =A(1,0,1) = A(1,1,0) =\omega ^k \notag \\
&A(1,2,0) = A(1,0,2)  =\omega ^{2k},
\end{align}
where $\omega = e^{2\pi i /3}$ and $k=0,1,2$ corresponds to three different  $\mathbb{Z}_3$ SPT phases, labeled as  $\text{SPT}^k$.
One can verify that the wavefunction is invariant under the $\mathbb{Z}_3$ symmetry, and the corresponding MPO for the symmetry is
 \begin{align}
\alpha = &| 00\rangle \langle 00 | +  | 01\rangle \langle 01 |+  | 02\rangle \langle 02 | +   \notag \\
                                    \omega^{2k} &| 10\rangle \langle 10 | +  | 11\rangle \langle 11 |+  | 12\rangle \langle 12 | +   \notag \\
                                   \omega^k &| 20\rangle \langle 20 | +  | 21\rangle \langle 21 |+  | 22\rangle \langle 22 |.
\end{align}

\begin{figure}[t]
\center{\epsfig{figure=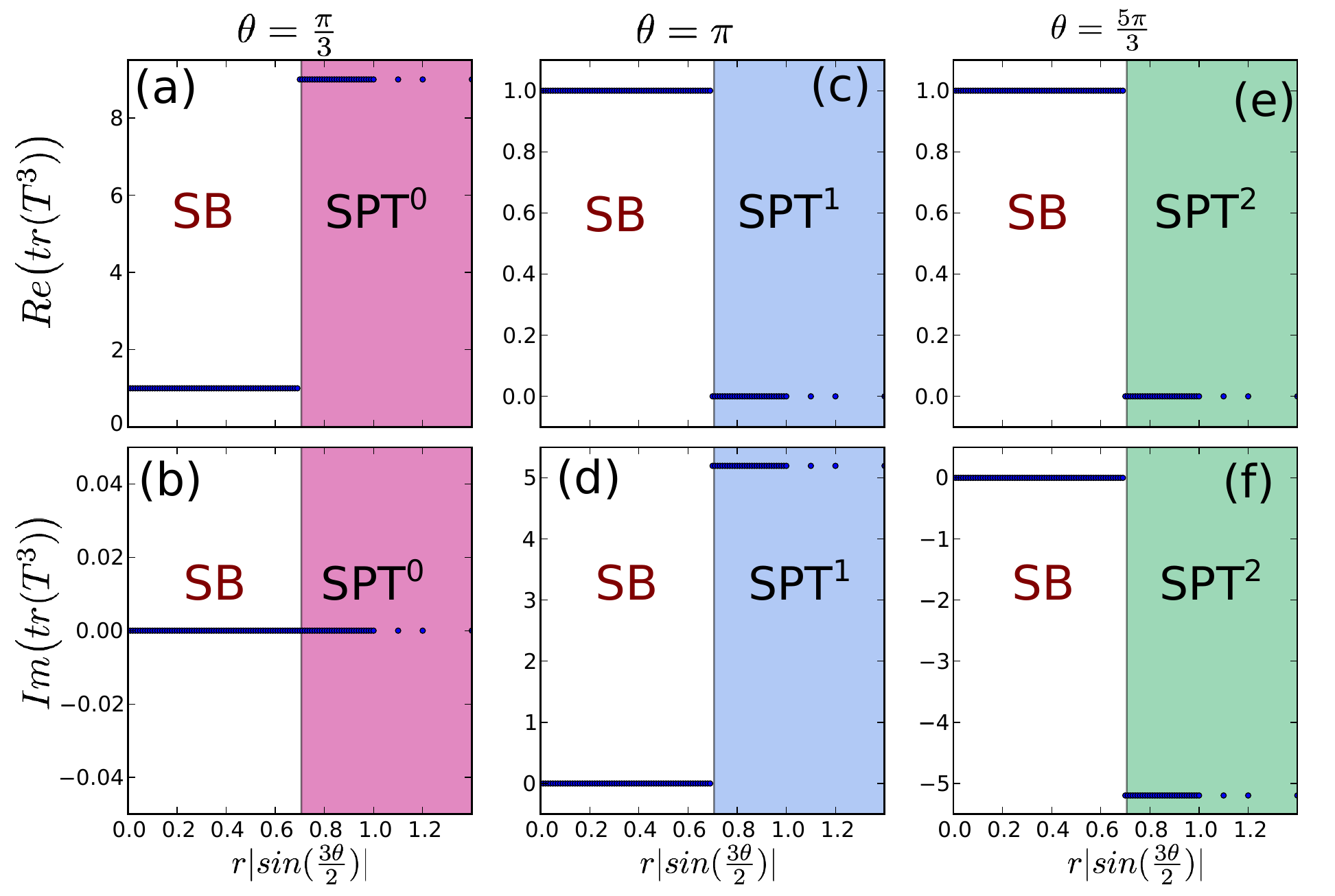,angle=0,width=8.5cm}}
\caption{ The $\mathbb{Z}_3$  SPT model represented by tensor [Eq.~(\ref{Z3deformed})]:  The real part of  the trace of  $T^3$ of  (a)  $\theta=\frac{\pi}{3}$ 
(c)  $\theta=\pi$, (e)  $\theta=\frac{4\pi}{3}$  and  the imaginary part of  the trace of  $T^3$ (b)  $\theta=\frac{\pi}{3}$ 
(d)  $\theta=\pi$, (f)  $\theta=\frac{4\pi}{3}$ as functions of parameter $r$ display a phase transition at critical point $r_c=0.708$  with $D_{cut}=18$ and 10 RG step.  }
  \label{fig:Z3_TRI}
\end{figure}

Similar to the $\mathbb{Z}_2$ case, in order to continuously connect all three SPT phases, we also consider a deformation on the above tensor preserving physical $\mathbb{Z}_3$ symmetry and the resulting tensor is
\begin{align}
\label{Z3deformed}
\tilde{A}^{(s_i,s_j,s_k)}_{(\alpha,\alpha',\beta,\beta',\gamma,\gamma')} =
  \begin{cases}
    1        \quad \text{if }s_i=s_j=s_k, \\
    r |\sin (\frac{3 \theta}{2})|   A^{k(\theta)}(s_i,s_j,s_k)   \ \text{otherwise},\\
  \end{cases}
\end{align}
where $0 \leq  \theta < 2\pi$ and $k(\theta)$ is a discrete label of the SPT sectors, given by the following step function,
\begin{align}
 k(\theta) =
  \begin{cases}
    0        \quad \text{if } 0 \leq  \theta < \frac{2\pi}{3} \\
    1        \quad \text{if }  \frac{2\pi}{3}  \leq  \theta < \frac{4\pi}{3} \\
    2        \quad \text{if }  \frac{4\pi}{3}  \leq  \theta < 2\pi \\ 
  \end{cases}.
\end{align}
We remark that one can also construct a two-parameter Hamiltonian such that the above family of states are the ground states, similar to the $\mathbb{Z}_2$ case. A natural question is that how are the three different $\mathbb{Z}_3$ SPT phases separated from one another? Is it by a region of symmetry-breaking phase? What is the phase diagram in terms of $r$ and $\theta$?

\begin{figure}[t]
\center{\epsfig{figure=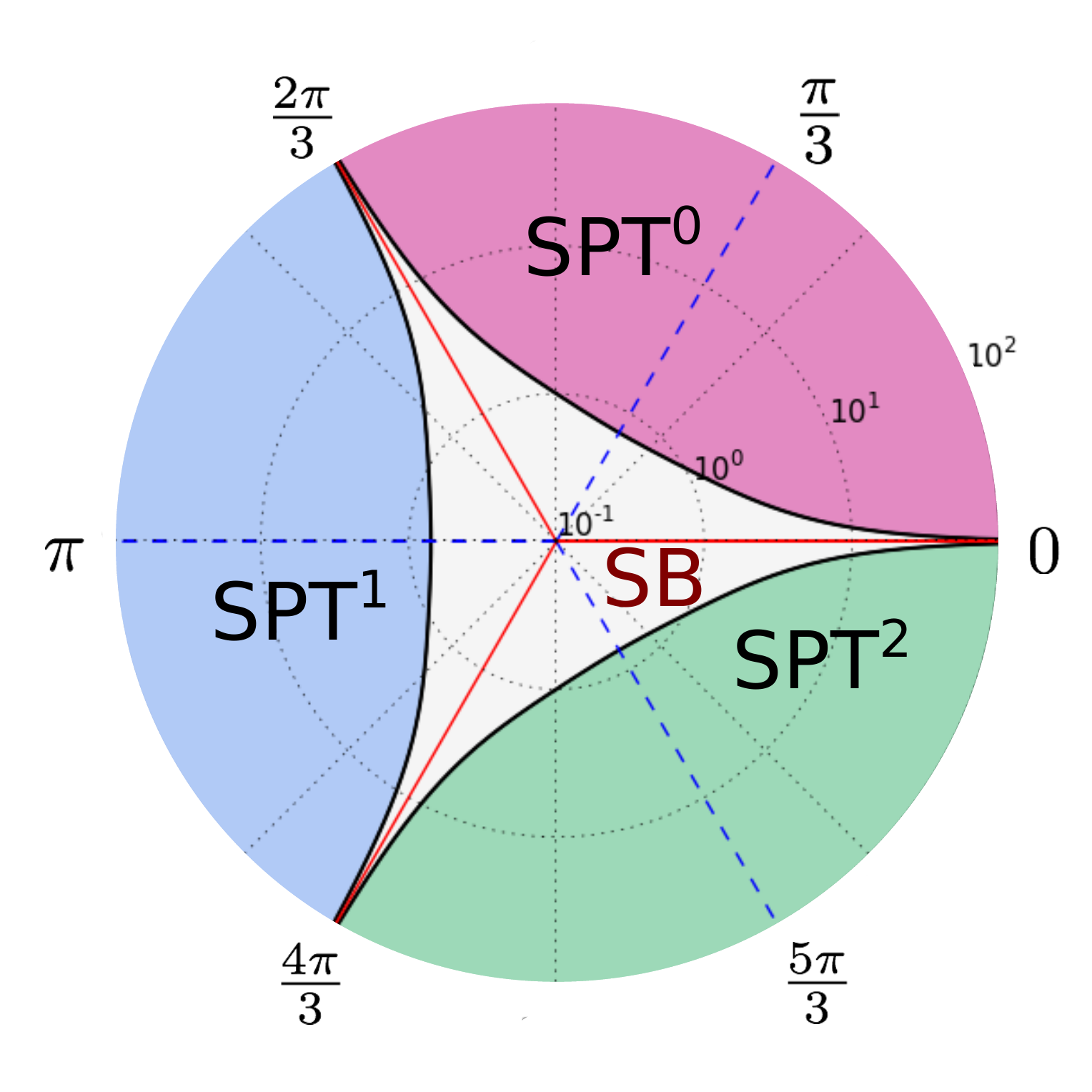,angle=0,width=6cm}}
\caption{Phase diagram of the deformed $\mathbb{Z}_3$  SPT model represented by  tensor [Eq.~(\ref{Z3deformed})]  with parameter $r$ and $\theta$. The radius of circle is $r$ and the logarithmic scale is used facilitate reading.
The phase diagram contain four different phases: the symmetry breaking phase (light grey region), $\text{SPT}^0$, $\text{SPT}^1$, and  $\text{SPT}^2$ phases.  The black line shows  the continuous phase transition.  The red lines of $\theta=0$, and $\theta=\frac{2\pi}{3}$ and $\theta=\frac{4\pi}{3}$ show exact symmetry breaking phase wave function.}
  \label{fig:Z3_SPT_PD}
\end{figure}

When $r|\sin(3\theta/2)|=1$, the tensor elements have the same weight and  these correspond to fixed-point $\mathbb{Z}_3$  SPT states in different sectors: $\text{SPT}^0$, $\text{SPT}^1$, and $\text{SPT}^2$, respectively.  
At $ \sin (3\theta/2)=0$, the tensor represents a symmetry-breaking state which is of equal weight superposition of  all $i$ ($i=0,1,2$), i.e., $|000\dots000\rangle+|111\dots111\rangle+|222\dots222\rangle$ (see the solid red lines in Fig.~\ref{fig:Z3_SPT_PD}). We expect that near these `branch cuts', there should be a spontaneous symmetry-breaking phase (to be labeled as SB). To find the region of the SB phase, we use HOTRG to compute the Landau local order parameter ${\cal O}\equiv\langle \delta_{s,0} \rangle-\frac{1}{3}$ vs. the parameter $r$ at a fixed value of $\theta=\pi/3$ and obtain the dependence as shown in Fig.~\ref{fig:Z3_TRI_MZ}. In particular, we find that the SB phase lies in the region $0\le r\le 0.708$ at $\theta=\pi/3$ and since the wavefunctions depend on the combination $r|\sin(3\theta/2)|$ we can conclude that the SB phase resides in the region $r|\sin(3\theta/2)|<0.708$. This seems to suggest that the three different SPT phases may be separated by this SB phase.

To explore the phase diagram, we employ our tnST approach to calculate the topological invariants and in particular $T^3$ to characterize possible  phases. 
First, we consider three angles on the phase diagram: (i) $\theta=\frac{\pi}{3}$ and $k(\theta)=0$,  (ii) $\theta=\pi$  and $k(\theta)=1$,  and (iii) $\theta=\frac{5\pi}{3}$ and $k(\theta)=2$. Our results are shown in Fig.~\ref{fig:Z3_TRI} by tuning $r$ but at these three different fixed values of $\theta$.
 When $0.708 \leq r <1.5$, the state belongs to the $\mathbb{Z}_3$  SPT phase, since we the modular  $T^3$ as follows:
 \begin{align}
T^3 =  \begin{pmatrix}
  1 & 0& 0 &0&0&0&0&0&0 \\
  0 & 1& 0 &0&0&0&0&0&0 \\
  0 & 0& 1 &0&0&0&0&0&0 \\
  0 & 0& 0 &\omega^k&0&0&0&0&0 \\
  0 & 0& 0 &0&\omega^k&0&0&0&0 \\
  0 & 0& 0 &0&0&\omega^k&0&0&0 \\
  0 & 0& 0 &0&0&0&\omega^k&0&0 \\
  0 & 0& 0 &0&0&0&0&\omega^k&0 \\
  0 & 0& 0 &0&0&0&0&0&\omega^k
 \end{pmatrix},
\end{align}
where $k \in \{0,1,2\}$ labels dinstinct SPT phases and $\omega= e^{2\pi i /3}$. The obtained modular matrices $S$ and $T$ agree with our analytical results in Sec.~\ref{sec:ST}. 
The three different  $\mathbb{Z}_3$  SPT phases are clearly characterized by the topological invariant $T^3$:
(i) for $\text{SPT}^0$ phase with $k=0$,  the trace of  $T^3$ is  $9$;  
(ii) for $\text{SPT}^1$ phase with $k=1$,  the trace of  $T^3$ is  $5.19615 i $;
(iii) for $\text{SPT}^2$ phase with $k=2$,  the trace of  $T^3$ is  $-5.19615 i $.
For $r\le 0.708$ at these three fixed angles,  we obtain trivial modular matrices and this detects the symmetry-breaking phase, which agrees with the spontaneous magnetization shown in Fig.~\ref{fig:Z3_TRI_MZ}.

\begin{figure}[t]
\center{\epsfig{figure=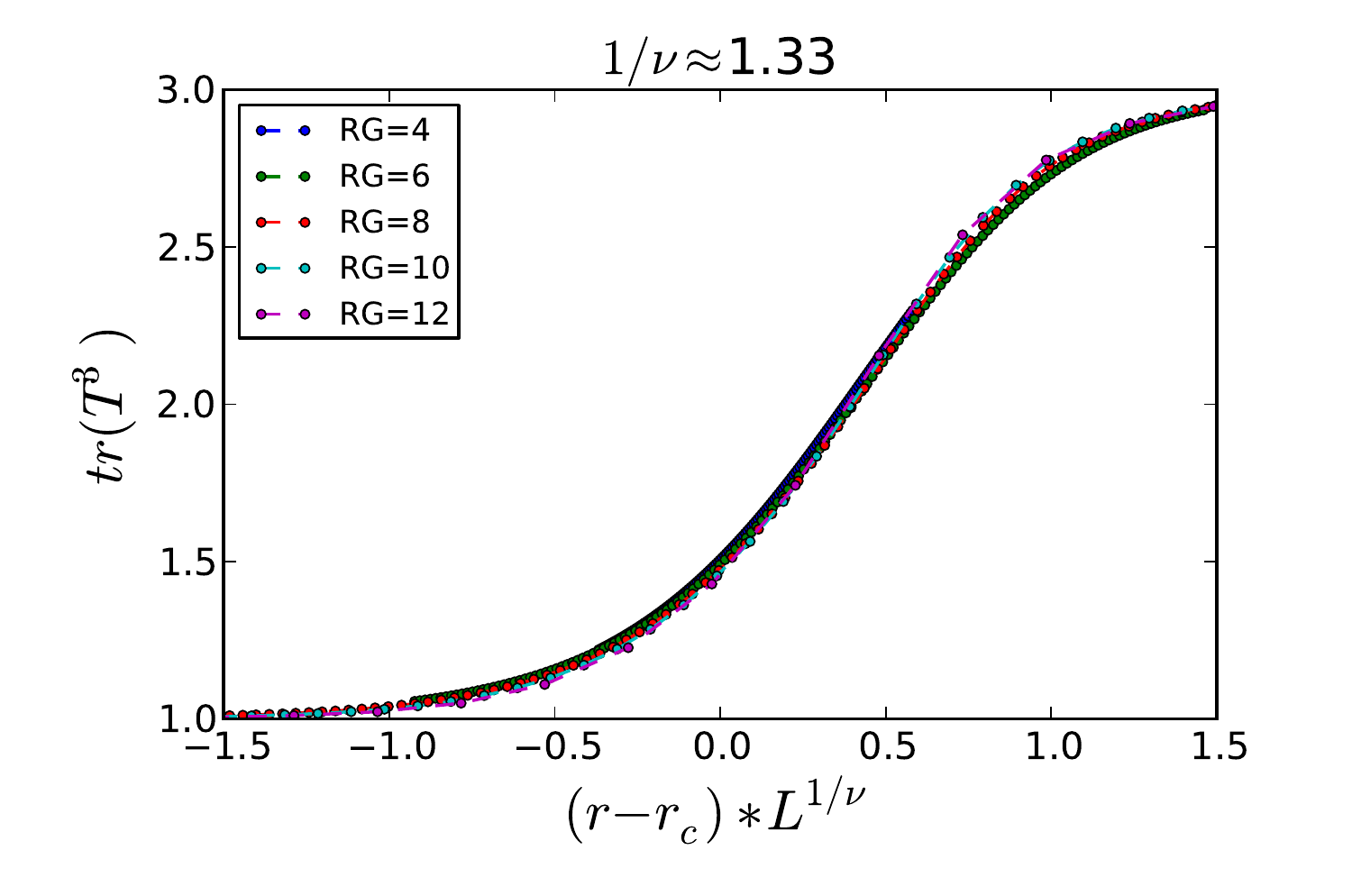,angle=0,width=8cm}}
\caption{ The trace of $T^3$ as a function $(r-r_c)\times L^{\nu}$ on  $\mathbb{Z}_3$  SPT model represented by  tensor [Eq.~(\ref{Z3deformed})]  with $\theta=\pi/3$.  }
  \label{fig:Z3_TRI_CP}
\end{figure}

After scanning over the parameter space, Eq.~(\ref{Z3deformed}), for the modular matrices via HOTRG, we obtain whole phase diagram parametrized by the radius $r$ and angle $\theta$ as shown in Fig.~\ref{fig:Z3_SPT_PD}. 
If the real part of  the trace of  $T^3$ is  $1$ and the imaginary part is $0$, we then identify it as in the symmetry breaking (SB) phase.   
If the real part of  the trace of  $T^3$ is  $9$ and the imaginary part is $0$, we then identify it as the    $\text{SPT}^0$ phase. 
If the real part of  the trace of  $T^3$ is  $0$ and the imaginary part is $5.19615$ or $-5.19615$, we the phase as  $\text{SPT}^1$ or $\text{SPT}^2$, respectively.  
As indicated in the phase diagram, the three solid red lines represent  branch cuts  and satisfy $|\sin(3\theta/2)|=0$.  The states on these three lines are all identical owing to the identical tensors by construction. Despite the branch cuts, the wavefunctions are smoothly connected along any curve between any two points on the phase diagram.
We determine the transition points from the change in the modular matrices and obtain the critical curves described by $r|\sin(3\theta/2)|=0.708$, as shown in Fig.~\ref{fig:Z3_SPT_PD}. That the transitions are continuous is consistent with the data collapse of the order parameter along the RG flow, as shown in Fig.~\ref{fig:Z3_TRI_CP}, with $\nu\approx 3/4$.
At fixed $\theta$ away from brach cuts, a phase transition occurs from the symmetry-break phase to one of the SPT phase, as $r$ increases from small to large values.  
At a fixed and large enough $r$ value, e.g., at $r=1$, by varying $\theta$ we find that the two SPT phases, e.g., $\text{SPT}^0$ and $\text{SPT}^1$, are separated by a region of the symmetry-breaking phase and there is no direct transition between the two at any finite $r$.

In the $\mathbb{Z}_2$ and $\mathbb{Z}_3$ examples investigated in this and previous subsections, the two SPT phases are separated by a symmetry-breaking phase and they do not make direct transition into each other. In the following, we construct another  $\mathbb{Z}_2$ SPT model that exhibits a direct transition between two SPT phases.

\subsection{ Phase transition directly between two SPT phases}
\label{subsec:twoSPT}
 
\begin{figure}[t]
\center{\epsfig{figure=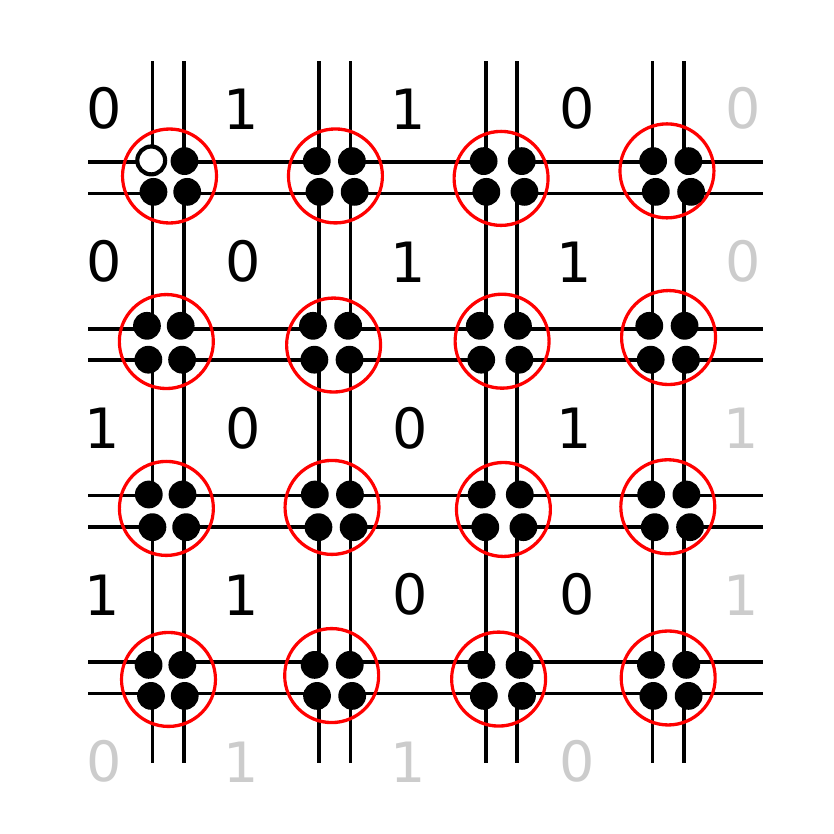,angle=0,width=5cm}}
\caption{One possible spin configuration of the symmetry breaking phase starting from the first site tensor $A[0,1,0,0]$. 
For simplicity, the labels around the red empty circle are the physical indices on each site. 
Other three ordered phases can be build up by starting from the first tensor $A[0,0,0,1]$, $A[1,0,1,1] $, and $ A[1,1,1,0] $.  }
\label{fig:SB_degenerate}
\end{figure}

\begin{figure}[ht]
\center{\epsfig{figure=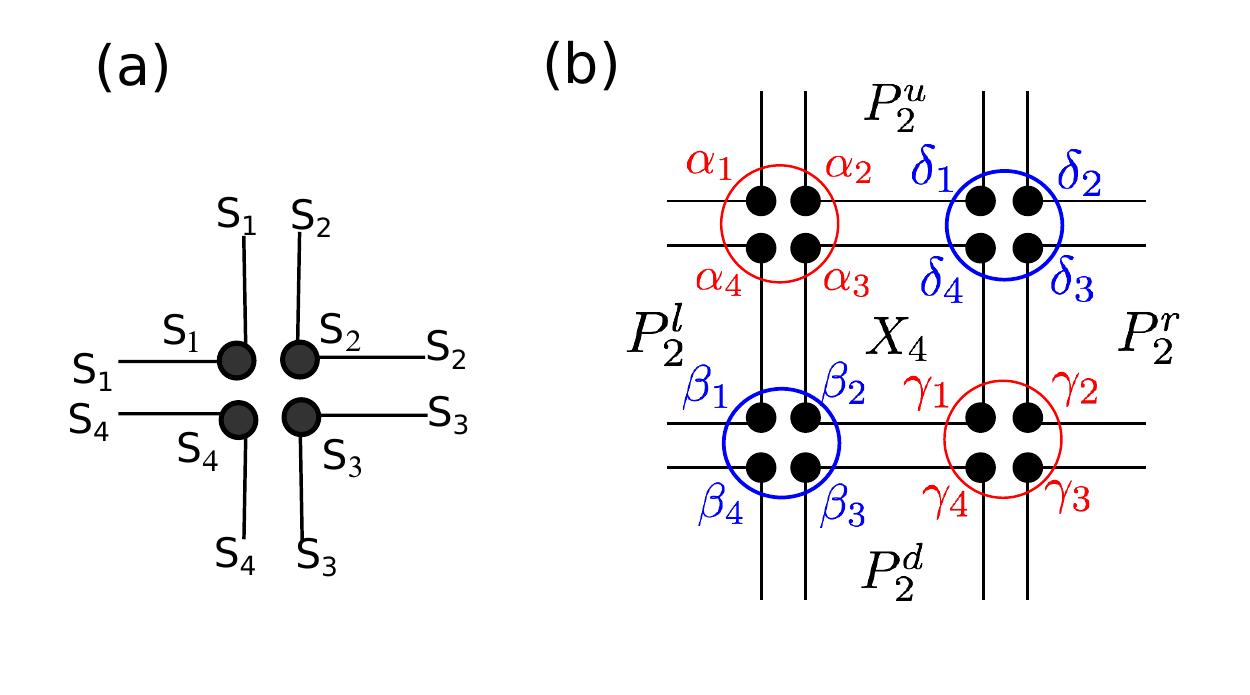,angle=0,width=7cm}}
\caption{(a) The tensor representation on the square lattice. The labels in circles are physical indices and the labels on the link are inner indices. (b) A local term in the Hamiltonian, which is tensor product of one four-site operator $X_4$ and four two-site operators $P_2$. }
\label{fig:square_local tensor}
\end{figure}

Here we construct another family of $\mathbb{Z}_2$ symmetric wavefunctions on the square lattice, with the local tensor $A[s_i,s_j,s_k,s_l]$ parametrized by $t$ [see Fig.~\ref{fig:square_local tensor}(a)], 
\begin{align}
\label{SPT_transition_tensor}
& A[0,0,0,0] = A[1,1,1,1] = A[0,0,1,1] = A[1,1,0,0] = 1  \notag \\
& A[1,0,0,1] = A[0,1,1,0] = A[0,1,0,1] = A[1,0,1,0] = 1  \notag \\
& A[0,0,1,0] = A[1,1,0,1] = A[1,0,0,0] = A[0,1,1,1] = 1  \notag \\
& A[0,1,0,0] = A[0,0,0,1] =  t  \notag \\
& A[1,0,1,1] = A[1,1,1,0] = |t|.
\end{align}
The parent Hamiltonian is listed in Eq.~(\ref{app:Z2SPTWF}) and its construction is explained in Appendix~\ref{sec:secondZ2}. The ground-state energy, by construction, is identically zero, and therefore, we expect any transition arising from this Hamiltonian and its ground state is continuous. 

The symmetry action is generated by the Pauli $\sigma_x$ on all four partons on all sites, and one can verify that the above wavefunction is indeed invariant under such a symmetry action.
As $t=1$, this tensor represents a fixed-point wavefunction of the trivial  $\mathbb{Z}_2$ SPT phase ($\text{SPT}^0$). 
As $t = -1$, this state is the fixed-point wavefunction of the non-trivial  $\mathbb{Z}_2$ SPT phase ($\text{SPT}^1$).  The wavefunctions above smoothly interpolate between the two phases. 

\begin{figure}[t]
\center{\epsfig{figure=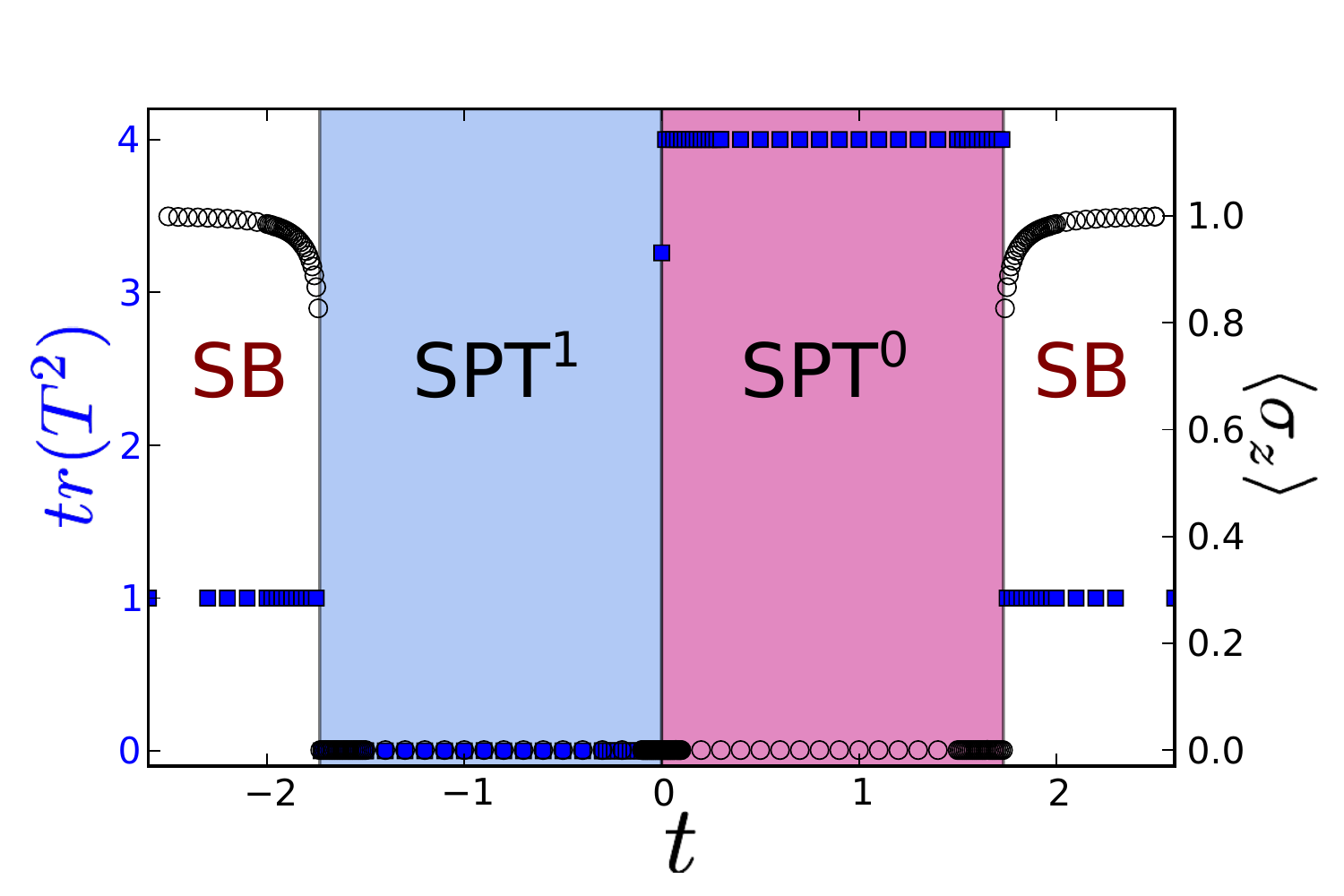,angle=0,width=8cm}}
\caption{  The $\mathbb{Z}_2$  SPT model represented by  the tensor  [Eq.~(\ref{SPT_transition_tensor})]: 
the trace of topological invariant matrices $T^2$ (solid blue square) and the magnetization $\langle \sigma_z \rangle$ at first site (empty black circle)  as functions of parameter $t$ display the  phase transitions with $D_{cut}=16$ and RG step $=10$. It shows the transition from symmetry breaking (SB) phase to non-trivial  $\mathbb{Z}_2$ symmetry protected topological order phase labeled as $\text{SPT}^1$ at $t_{c1} \approx -1.7321$ and from  $\text{SPT}^1$ phases  to trivial $\mathbb{Z}_2$ symmetry protected topological order phase labeled as $\text{SPT}^0$  at $t_{c2}=0$ and from $\text{SPT}^0$ phase  to symmetry breaking phase at   $t_{c3} \approx 1.7321$. }
\label{fig:WF_one}
\end{figure}

For  large magnitudes of $t$, the tensor in Eq.~(\ref{SPT_transition_tensor}) has four dominant local physical configurations: $ A[0,1,0,0] = A[0,0,0,1] =  t  ;\  A[1,0,1,1] = A[1,1,1,0] = |t| $.
Obviously, this shows an ordered phase with a symmetry-breaking pattern of a $4\times 4$ unit cell, represented by the effective degree of freedom on each plaquette as shown in Fig.~\ref{fig:SB_degenerate}. 
This gives rise to four-fold symmetry breaking of the ground state. 
With the symmetry breaking pattern understood we can construct local order parameters to characterize this phase. 
 By computing the $\langle \sigma_z \rangle$ of the parton on the top-left corner at the first site (labeled as a blue empty circle in Fig.~\ref{fig:SB_degenerate}) and averaging it over all $4\times4$ unit cells we obtain the behavior of the order parameter for this symmetry-break phase, shown in Fig.~\ref{fig:WF_one}, and find quantum phase transitions at $|t|\approx 1.73$.

Given the understanding of symmetry-break phases at large $|t|$ and the existence of SPT phases around $|t|=1$, a natural question is whether there is an additional phase in between the two SPT phases? To answer this we shall calculate the modular matrices for a wide range of t. But we need to first examine the MPO's related to the symmetry action. The inner symmetry operator for the wavefunction in Eq.~(\ref{SPT_transition_tensor}) is $ X \otimes X \alpha$ and either $\alpha = | 00\rangle \langle 00|+ | 01\rangle \langle 01|+| 10\rangle \langle 10|+| 11\rangle \langle 11|$ for $t\ge 0$ or $\alpha = | 00\rangle \langle 00|+ i| 01\rangle \langle 01|+i|10\rangle \langle 10|+| 11\rangle \langle 11|$ for $t\le 0$.  Then we use the HOTRG to evaluate the wavefunction overlaps for the modular matrices by inserting various combinations of MPO's. 
Let us emphasize that when we implement our scheme we numerically check which MPO can make the wavefunction invariant and then take that MPO for the evaluation of modular matrices (even though we know which one should be).

Our results  are shown in Fig.~\ref{fig:WF_one}.
First, we see that when $|t|\geq1.7321$, all components of  $4 \times 4$ topological invariant  $T^2$  matrices are $0$ except for  $T^2 [1,1]=1$. 
We can thus deduce that these two regions are a symmetry-break phase.
We find that the SPT phases are in the regions $0<|t|<1.7321$  since we obtain nontrivial  $T^2$ matrices, same as Eq.~(\ref{T2matrixk0}) for $t>0$ and Eq.~(\ref{T2matrixk1}) for $t<0$.
Thus a quantum phase transition separates these two SPT phases at $t=0$. Is the transition is discontinuous or continuous?

\begin{figure}[t]
\center{\epsfig{figure=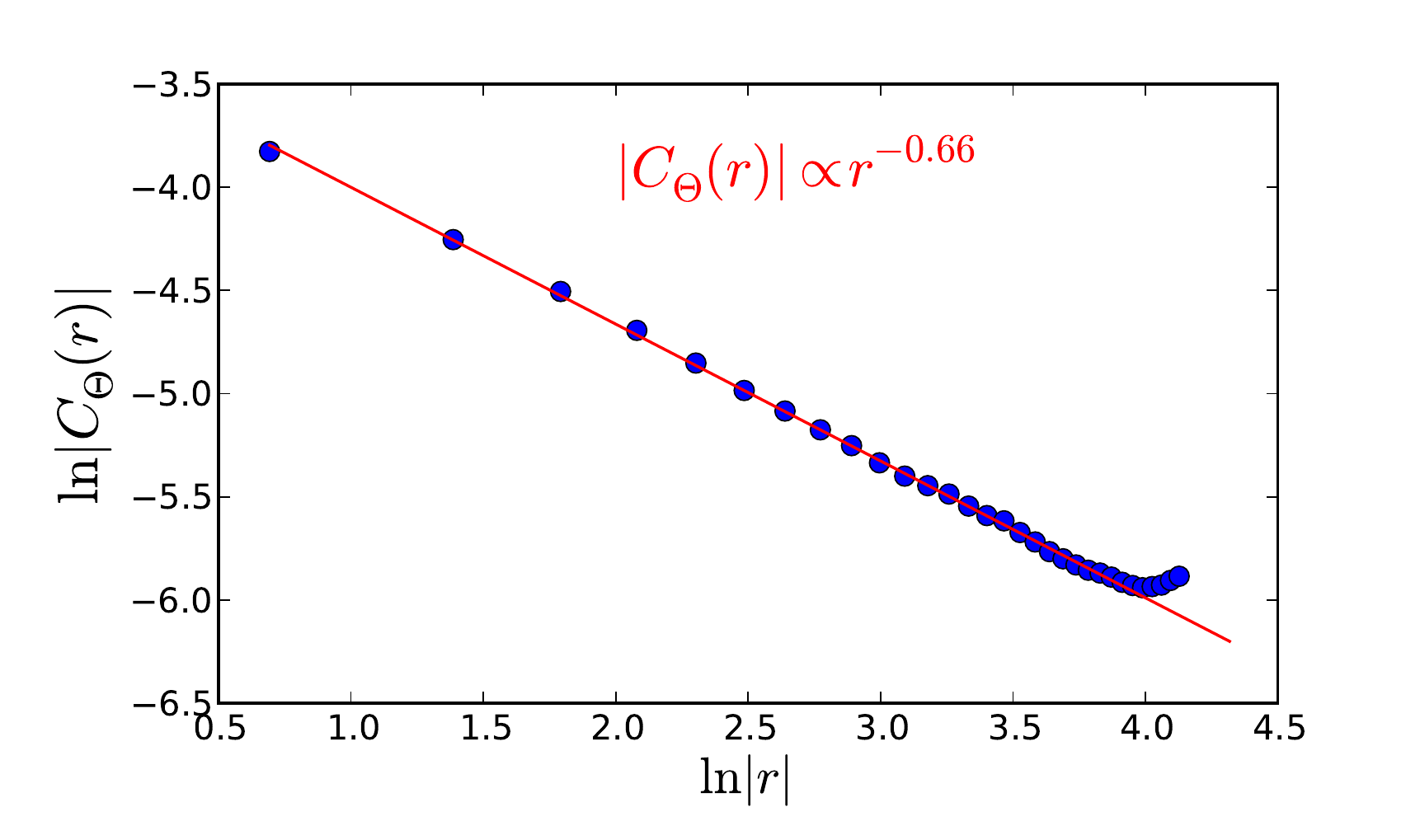,angle=0,width=8cm}}
\caption{The correlation function at $t=0$ as a function of the distance $r=|R_i-R_j|$. } 
 \label{fig:wfone_correlation}
\end{figure}

We need further evidence to determine what the transition is. 
The relevant quantity of interest is the correlation function which can distinguish whether the transition point is critical or not, and 
it is defined as,
\begin{align}
C_{\Theta} ( |R_i-R_j| )= \langle \Theta(R_i) \Theta(R_i)  \rangle -  \langle \Theta(R_i) \rangle  \langle \Theta(R_i)  \rangle,
\end{align}
where the observable is chosen to be
\begin{align}
\Theta(R_i)  = |  \begin{smallmatrix} 0&0 \\ 0 &0 \end{smallmatrix}   \rangle   \langle   \begin{smallmatrix} 0&0 \\ 0 &0 \end{smallmatrix} |
- |  \begin{smallmatrix} 1&1 \\ 1 &1 \end{smallmatrix}   \rangle   \langle   \begin{smallmatrix} 1&1 \\ 1 &1 \end{smallmatrix} |,
\end{align}
namely, $\Theta(R_i) $ counts 1 (or -1) if the local state $ |\begin{smallmatrix} 0&0 \\ 0 &0 \end{smallmatrix}   \rangle $
 (or $ |\begin{smallmatrix} 1&1 \\ 1 &1 \end{smallmatrix}   \rangle $) on the site $R_i$ in a given configuration, and otherwise $0$. 
In our tensor renormalization calculations, the correlation function at $t=0$ is evaluated on the square lattice with size being $128 \times 128$. 
The distance $ |R_i-R_j| $ is chosen along the horizontal direction.
The results with bond dimension $D_{cut}=32$ show that the correlation function  is algebraically decaying, as shown in Fig.~\ref{fig:wfone_correlation} and the anomalous exponent  $\eta\approx 0.66$. This shows that $t=0$ is a critical point   and the transition is continuous.

\section{conclusion} \label{sec:conclusion}

We have described a  scheme to evaluate the modular matrices for investigating SPT phases. This  tnST scheme can also apply to topological phases and can be used to detect symmetry-breaking phases as well. We have demonstrated by various model studies that our scheme can be implemented using tensor-network methods and can accurately characterize and identify these different phases and related phase transitions. Thus modular matrices  provide a unifying framework for characterizing and identifying gapped phases and can be used as a theoretical order parameter. For the spontaneous symmetry-breaking phases occurring in our constructions, we have additionally characterized them by evaluating the traditional Landau-type local order parameters, and their behavior agrees with that of modular matrices. The numerical evaluations of modular matrices involve overlaps of wavefunctions under symmetry twists, and we have conveniently used the higher-order tensor renormalization group for the actual numerical implementations but any other contraction algorithm can be used. 

In addition to classification of SPT phases, recently, the nature of the phase transition between two SPT phases has also been investigated~\cite{Tsui2015330,ChenNPB}. In particular, Tsui et al.~\cite{Tsui2015330} proposed that there are three scenarios regarding the transition between one trivial and another nontrivial SPT phases, if the nontrivial SPT phase satisfies the so-called non-double-stacking condition: (i) a direct continuous transition; (ii) a direct discontinuous transition; (iii) an intermediate spontaneous symmetry-breaking phase separating the two SPT phases.
The nontrivial $\mathbb{Z}_2$ SPT phase indeed satisfies this, and we have constructed two $\mathbb{Z}_2$ models, where we know analytically both their Hamiltonians and their ground states. For the first one model, the two $\mathbb{Z}_2$ are separated by a spontaneous symmetry-breaking phase. This corresponds to the scenario (iii), where it was argued that  there is a corresponding 3D SPT with symmetry group being $\mathbb{Z}_2\times \mathbb{Z}_2^T$ but either $\mathbb{Z}_2$ symmetry or both $\mathbb{Z}_2$ and $\mathbb{Z}_2^T$ are  spontaneously broken.  For the second model we constructed, we found that the two SPT phases have a direct transition into each other and the quantum phase transition is continuous, realizing the scenario (i). According to Ref.~\cite{Tsui2015330}, this corresponds to the case where the corresponding 3D SPT phase has gapless boundary, but can be gapped out by introducing a $\mathbb{Z}_2^T$ breaking field.  
We have remarked earlier (also in the Appendix) that as our construction is wavefunction-based and the parent Hamiltonians are non-negative operators, the transitions arising from our models will be continuous. But it was realized that a Hamiltonian-based construction, e.g., a linear interpolation from the two parent Hamiltonians ($H_0$ and $H_1$) corresponding to two SPT fixed-point states, $\tilde{H}(\lambda)=(1-\lambda) H_0 + \lambda H_1$, a first order transition can occur~\cite{Siddhardh14,Tsui2015330}. The nontrivial $\mathbb{Z}_3$ SPT phases do not satisfy the non-double-stackable condition, so the possible transition scenarios are not as clear. But the $\mathbb{Z}_3$ model we constructed has transitions between any two of the SPT phases separated by a common spontaneous symmetry-breaking phase, i.e., the scenario (iii).

As we have remarked earlier, a full-fledge application would involve (i) solving the Hamiltonian and obtaining the ground state wavefunction(s), (ii) identifying the symmetry and the associated matrix-product operator, and then (iii) employing our tnST approach here to obtain the modular matrices. Regarding point (i), there have been various TN algorithms for finding ground states from Hamiltonians~\cite{VerstraeteMurgCirac,OrusTN}. Regarding (ii), there have been in-depth theoretical investigation of MPO's in topological and symmetry-protected topological order~\cite{VerstraeteTO_MPO_2014,VerstraeteSPT_MPO_2014} and it would be useful to develop a numerical scheme to find these symmetry-related MPO's. Together with what we have demonstrated here, it should be feasible to have these pieces put together and obtain a useful numerical toolbox to investigate gapped phases with tensor-network methods.  

Even though we have limited our discussions to mostly $\mathbb{Z}_N$ symmetry (or gauge symmetry for topological order), our method applies to any other finite group, either Abelian or non-Abelian. It will be interesting to include anti-unitary symmetry elements, such as the time reversal~\cite{GaugingTime} in the same framework. Moreover, since our approach uses wavefunction overlaps, it should be possible to extend it to three dimensional intrinsic topological and symmetry-protected topological phases, as well as symmetry-breaking ones.   We leave these for future investigation.

\section*{Acknowledgements}

The authors would like to thank Lukasz Fidkowski, Janet Hung and Abhishodh Prakash   for useful discussions.
This work was supported by the National Science Foundation under
Grants No. PHY 1314748 and No. PHY 1333903.

\appendix

\section{Hamiltonian of the $\mathbb{Z}_2$ model in Sec.~\ref{subsec:ZNSPT}}
\label{sec:firstZ2}
Here we construct the parent Hamiltonian for the $\mathbb{Z}_2$ family of wavefunctions in Eq.~(\ref{eqn:Z2WF}) that was described in Sec.~\ref{subsec:ZNSPT}. 
The system resides on the hexagonal lattice;  see Fig.~\ref{fig:ZN_tensor}(b)
At $g=1$, the tensor product state is a fixed-point $\mathbb{Z}_2$ $\text{SPT}^{0}$ wavefuntion and consists of   a product of all plaquette states, $|\psi \rangle_p = | 000000\rangle +|111111\rangle $, each composed of six spins or partons from six neighboring sites around a plaquette labeled by $p$. 
The parent Hamiltonian of the wavefunction at $g=1$ (such that the latter is the ground state of the former) is thus constructed as a sum of local  projector terms around each plaquette $\{p\}$, $H_0=\sum_p h_{0}^{\{p\}}$,
where $h_{0}^{\{p\}}$ acts on the six spins on the plaquette and on the twelve spins in the neighboring  one-third  plaquettes,
\begin{align}
h_{0}^{\{p\}} =& ( \mathbb{I}- X_6)   \otimes  ( \mathbb{I}-P_2^a) \otimes ( \mathbb{I}-P_2^b) \otimes ( \mathbb{I}-P_2^c) \notag \\
                       &                   \otimes ( \mathbb{I}-P_2^d )  \otimes ( \mathbb{I}-P_2^e )  \otimes ( \mathbb{I}-P_2^f).
\end{align}
In the above, $X_6$ is a six-site operator $|000000\rangle  \langle 111111|+|111111\rangle \langle 000000|$ and acts on the six spins on the plaquette $p$.
The $P_2$ is a projection operator  $|00\rangle  \langle 00|+|11\rangle \langle 11|$ that acts  on the two spins in  every neighboring one-third plaquette.

For the  fixed-point $\mathbb{Z}_2$ $\text{SPT}^{1}$ wavefuntion (i.e., at $g=-1$),  the Hamiltonian is of a similar form, except with an additional phase twist,  
\begin{align}
 H_1=\sum_p h_{1}^{\{p\}} = \sum_p  \mathbb{U}_p   h_{0}^{\{p\}}  \mathbb{U}_p ^{\dagger},
\end{align}
where  $  \mathbb{U}_p =  U_L^{[\alpha]} U_L^{[\beta]}  U_L^{[\gamma]} U_L^{[\delta]} U_L^{[\eta]} U_L^{[\lambda]} $ acts on the six sites around the plaquette and gives a phase twist depending on the spin configurations, and 
\begin{align}
U_L^{[\alpha]} = \mathbb{I}-  2 (| 110 \rangle \langle 110 | + | 101 \rangle \langle 101 |+| 011 \rangle \langle 011 | )
\end{align}
acts on all three partons  $|\alpha_1\alpha_2 \alpha_3 \rangle $  on  site $\alpha$.

Since the wavefunctions are connected smoothly by the deformation $Q(g)$ defined in Eq.~(\ref{eqn:DeformationQ}), we can now easily construct their parent Hamiltonian, with the  ground states being the wavefunctions in Eq.~(\ref{eq:ZNTPS}),  
\begin{equation}
\label{app:Z2WF}
H(g) \equiv\begin{cases}
    \sum_p    \mathbb{R}_p(g)   \  h_1^{\{p\}} \  \mathbb{R}_p(g)    & \quad \text{if } g<0,\\
    \sum_p    \mathbb{R}_p(g)  \ h_0^{\{p\}}  \ \mathbb{R}_p(g)     & \quad  \text{if } g\ge 0,\\
  \end{cases}
\end{equation}
where  $  \mathbb{R}_p(g) =  R^{[\alpha]}(g) \ R^{[\beta]}(g) \ R^{[\gamma]}(g) \ R^{[\delta]}(g) \ R^{[\eta]}(g) \ R^{[\lambda]}(g) $ acts on the six sites around one plaquette and $R(g) = |g|Q^{-1}(g)$ and $Q^{-1}(g)$ is regularized by $|g|$ when $g\rightarrow 0$.  One can verify that the two branches are smoothly connected, i.e., $H(0^{+})=H(0^{-})$, so $H(g)$ is a continuous Hamiltonian in terms of the parameter $g$. 

By construction, the ground-state energy of the Hamiltonian is always zero, as it consists of non-negative operators, and hence one expects that the transitions appearing in this model have to be continuous. Our construction here can be regarded as the two-dimensional generalization of those quantum phase transitions in matrix product systems~\cite{WolfOrtizVerstraeteCirac}. In contrast, the Hamiltonian constructed by interpolating the two, i.e., $\tilde{H}(\lambda)\equiv (1-\lambda) H_0 + \lambda H_1$, might have different order of transitions, such as first-order constructed in Refs.~\cite{Siddhardh14} and~\cite{Tsui2015330}.

\section{Hamiltonian of the $\mathbb{Z}_2$ model in Sec.~\ref{subsec:twoSPT}}
\label{sec:secondZ2}

Here we consider another family of $\mathbb{Z}_2$ symmetric wavefunctions in Eq.~(\ref{SPT_transition_tensor})  and  construct a one-parameter  parent Hamiltonians (parameterized by $t$) such that these wavefunctions are the unique ground states (except at transitions). 
As $t=1$, this tensor represents a fixed-point wavefunction of the trivial  $\mathbb{Z}_2$ SPT phase ($\text{SPT}^0$). 
In the ground state,  fours parton spins, each from four neighboring sites,  are entangled as $|\psi \rangle = |0000 \rangle +|1111 \rangle$, and the total wavefunction is a product of all plaquette wave function. 
The Hamiltonian for which this wavefunction is the unique ground state is constructed as a sum of local terms around each plaquette, $H_0=\sum_p h_{0}^{\{p\}} $, where
 $h_{0}^{\{p\}} $, as illustrated in Fig.~\ref{fig:square_local tensor}(b), acts on the four spins on each plaquette (with each spin belonging to a neighboring site around the plaquette) and on the eight spins on the four neighboring half plaquettes,
\begin{align}
 \label{eq:z2spt_h0}
h_{0}^{\{p\}}  =   (\mathbb{I}\!-\! X_4)\otimes  (\mathbb{I}\!-\!  P_2^u) \otimes  (\mathbb{I}\!-\!P_2^d) \otimes  (\mathbb{I}\!-\!P_2^l) \otimes ( \mathbb{I}\!-\!P_2^r).
\end{align}
In the above, $X_4$ is a four-site operator $|0000\rangle  \langle 1111|+|1111\rangle \langle 0000|$ and acts on the four spins on the plaquette $p$.
The projector $P_2=|00\rangle  \langle 00|+|11\rangle \langle 11|$  acts  on the two spins in  every neighboring half plaquette.  
This is equivalent to the CZX Hamiltonian~\cite{Chen_2DCZX_2011}, but the symmetry there is a more complicated $\mathbb{Z}_2$ operation; see Appendix~\ref{app:CZX}.

As $ t= -1$, this  is the fixed-point wavefunction of the non-trivial  $\mathbb{Z}_2$ SPT phase ($\text{SPT}^1$). 
The  parent Hamiltonian of this ground-state wavefunction is constructed, similarly, as a sum of local terms around each plaquette,
 \begin{align}
 \label{eq:z2spt_h1}
 H_1=\sum_p h_{1}^{\{p\}}  = \sum_p  \mathbb{U}_p   h_{0} ^{\{p\}}   \mathbb{U}_p ^{\dagger}
\end{align}
where  $  \mathbb{U}_p =  U_L^{[\alpha]} U_L^{[\beta]}  U_L^{[\gamma]} U_L^{[\delta]} $, similar to that in the previous section,  acts on the four sites around one plaquette as shown in Fig.~\ref{fig:square_local tensor}(b), and 
\begin{align}
U_ L^{[\alpha]} = \mathbb{I}-  2 (  |  \begin{smallmatrix} 0&1 \\ 0 &0 \end{smallmatrix}   \rangle   \langle   \begin{smallmatrix} 0&1 \\ 0 &0 \end{smallmatrix} |
+|  \begin{smallmatrix} 0&0 \\ 0 &1 \end{smallmatrix}   \rangle   \langle   \begin{smallmatrix} 0&0 \\ 0 &1 \end{smallmatrix} | )
\end{align}
acts on all four partons   $ |  \begin{smallmatrix} \alpha_1 &\alpha_2 \\ \alpha_4 &\alpha_3 \end{smallmatrix}   \rangle $ on the site $\alpha$. 

The parent Hamiltonian of the one-parameter ground state represented by the tensor product state in Eq.~(\ref{SPT_transition_tensor}) can thus be constructed  as
\begin{widetext}
\begin{equation}
\label{app:Z2SPTWF}
H(t) \equiv\begin{cases}
 \sum_p    \big[ R^{[\alpha]}(t)  \otimes  R^{[\beta]}(t) \otimes R^{[\gamma]}(t) \otimes R^{[\delta]}(t) \big] \  h_{1}^{\{p\}}   \ \big[ R^{[\alpha]}(t)  \otimes  R^{[\beta]}(t) \otimes R^{[\gamma]}(t) \otimes R^{[\delta]}(t) \big]       & \quad \text{if } t<0,\\
 \sum_p   \big[ R^{[\alpha]}(t)  \otimes  R^{[\beta]}(t) \otimes R^{[\gamma]}(t) \otimes R^{[\delta]}(t) \big]      \  h_{0}^{\{p\}}     \
   \big[ R^{[\alpha]}(t)  \otimes  R^{[\beta]}(t) \otimes R^{[\gamma]}(t) \otimes R^{[\delta]}(t) \big]         & \quad  \text{if } t\ge 0,\\
  \end{cases}
\end{equation}
\end{widetext}
 where  $h_0^{\{p\}} $ and $h_1^{\{p\}} $ are described in  Eqs.~(\ref{eq:z2spt_h0}) and~(\ref{eq:z2spt_h1}), respectively, and $R(t) = |t|Q^{-1}(t) $ and  $Q(t)$ is the local deformation operator acting on all four partons on each sites as shown in Fig.~\ref{fig:square_local tensor}(a), with 
 \begin{equation}
Q(t) = \sum_{s_1 s_2 s_3 s_4=0}^1    q_{s_1 s_2 s_3 s_4}     |  \begin{smallmatrix} s_1&s_2 \\ s_4 &s_3 \end{smallmatrix}   \rangle   \langle   \begin{smallmatrix} s_1&s_2 \\ s_4 &s_3\end{smallmatrix} |,  \notag \\
  \end{equation}
where  
  \begin{equation}
    q_{s_1 s_2 s_3 s_4} =  \begin{cases}
    |t| & \ \text{if }   (s_1 s_2 s_3 s_4 )=0100,0001,1011, 1110,   \\
    1 & \ \text{otherwise}.    \\
   \end{cases}
 \end{equation}

One can verity that $H(0^{+})=H(0^{-})$ and hence the $H(t)$ is  continuous in $t$.  By construction, the ground states have zero energy, as the Hamiltonian consists of non-negative operators. Any transition exhibited in the Hamiltonian~(\ref{app:Z2SPTWF}) thus has to be continuous.

\section{ The $\mathbb{Z}_2$ topological model with string tension on hexagonal lattice} \label{app:Z2_state_hex}
In this Appendix and the following ones, we present additional models of exhibiting intrinsic and symmetry-protected topological order. 
 
Consider the toric code model on the hexagonal lattice. 
The ground state wave function is still a equal wight superposition of string-net configurations and it has a simple  tensor product representation by a rank-three tensor $P_{\alpha,\beta,\gamma}$ with three internal indices running over $0$ and $1$ on the vertex and a rank-three tensor $G^s_{\alpha,\beta}$ with one physical index $s$ running over the spin-$1/2$ basis states on the link. 
The wavefunction is then given by~\cite{Chen_Gu_Z2string_2010} 
\begin{align}
|\Psi \rangle = \sum_{s_i} tTr (\otimes_v P \otimes_l G^{s_i})|s_1,s_2,....\rangle, 
\end{align}
where $v$ labels vertices and $l$ links. Specifically, 
\begin{align}
 &P_{000}=P_{011}=P_{101}=P_{110}=1,
 \end{align}
and 
\begin{align}
& G^0_{00}=G^1_{11} = 1,
 \end{align}
where we see that the rank-three tensor $G$ behaves like a projector that identifies the physical and internal bond indices. Then we consider a deformation $Q(g)=\sqrt{g} |0 \rangle \langle 0| + |1 \rangle \langle 1|$ and $g \geq0$, which applies to the physical indices, $|\Psi _{(g)} \rangle  = Q(g) \otimes Q(g) \otimes...\otimes Q(g) |\Psi \rangle $.
To obtain a TPS associated with the site tensor, we can split every spin-$1/2$ particle into two and associate each vertex with three spins.
Then the deformed wavefunctions  have a simple TPS by a single local tensor $A^{s_i,s_j,s_k}_{\alpha,\beta,\gamma}$,
\begin{align}
\label{Z2_TOhex}
&A^{000}_{000} = \sqrt{g}, \ A^{011}_{011}=  A^{101}_{101}=A^{110}_{110}=1 \notag\\
& \text{all other terms are zero}.
 \end{align}
This wavefunction is a  $\mathbb{Z}_2$ gauge-symmetric state with string tension. 
As $g=1$, the state is exact  $\mathbb{Z}_2$ toric-code state. As $g\gg 1$, the tensor approaches a product state of all  $0$'s: $|00\dots0\rangle$.
By tuning the parameter $g$, a transition must occur. 

The local gauge transformation of this  $\mathbb{Z}_2$ model is  
$Z=\bigl(\begin{smallmatrix}
1&0 \\ 0 &-1
\end{smallmatrix} \bigr)$ acting on the bond degrees of freedom.
By applying various combinations of gauge transformation in two orthogonal directions to the wavefunction, we can determine the modular matrices. 
As we increase $g$, the state will go through a phase transition from a topological phase to a non-topological phase, and in Ref.~\cite{Chen_Gu_Z2string_2010},  it was shown that a quantum critical  point occurs at $g_c \approx 1.75$ by calculating the topological entanglement entropy, which agrees with the exact transition point, $g = \sqrt 3 = 1.73205$, via a mapping from the wavefunction norm square to the partition function of the classical Ising model and given by $\beta = \ln g /2$.

\begin{figure}[t]
\center{\epsfig{figure=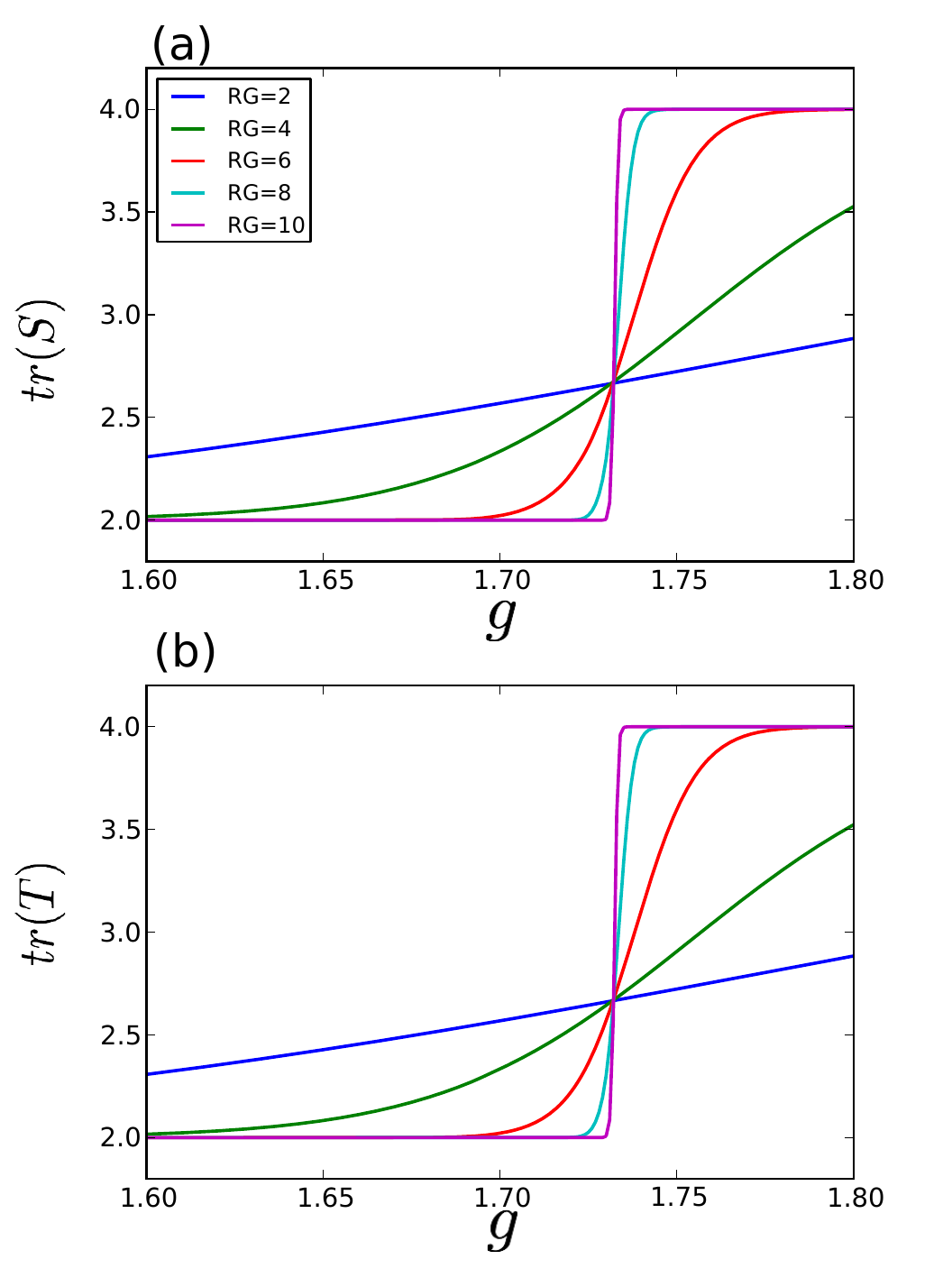,angle=0,width=8cm}}
\caption{ The  $\mathbb{Z}_2$ topological model represented by tensor~(\ref{Z2_TOhex}) on the hexagon lattice:  the trace of modular matrices (a) $S$ and  (b) $T$ as functions of parameter $g$  display a phase transition at critical point $g_c \approx 1.732$ under the renormalization flow.} 
 \label{fig:Z2_hex}
\end{figure}

We use our scheme to evaluate the $S$ and $T$ matrices and the results are shown in Fig.~\ref{fig:Z2_hex}, wihch detect a quantum phase transition around $g_c \approx 1.732 $,  separating the topological phase characterized by $tr(S)=tr(T)=2$ and the non-topological phase characterized by $tr(S)=tr(T)=4$.

\begin{figure}[t]
\center{\epsfig{figure=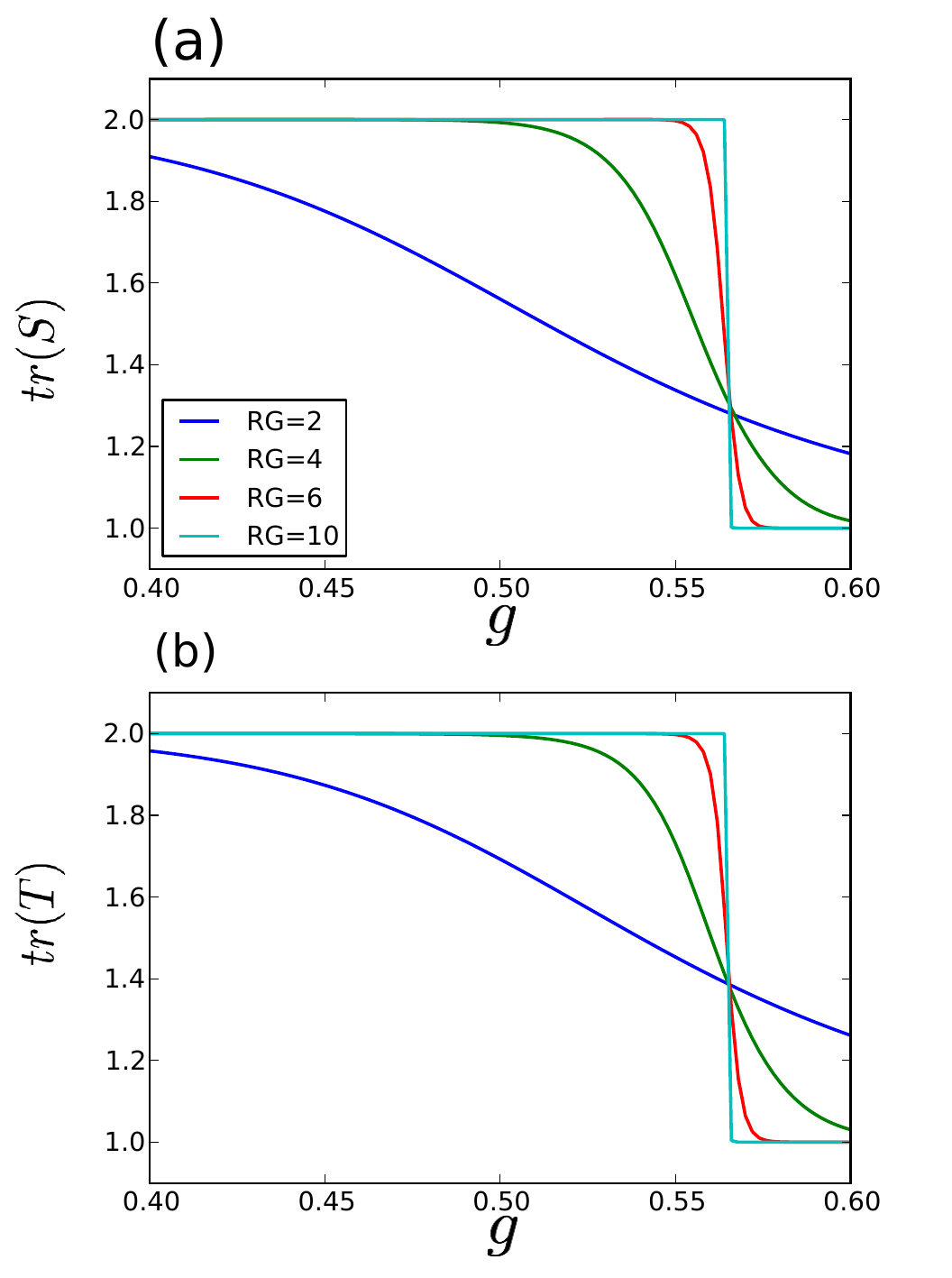,angle=0,width=8cm}}
\caption{  The $\mathbb{Z}_2$ topological model on the square lattice:  the trace of modular matrices (a) $S$ and  (b) $T$ as functions of parameter  display a phase transition at critical point $\theta_c \approx 0.56$  under the renormalization flow.} 
 \label{fig:Z2string}
\end{figure}

\section{ The $\mathbb{Z}_2$ topological model with string tension on the square lattice} \label{app:Z2_state_square}

Here we consider another $\mathbb{Z}_2$ toric-code model under a string tension $g$, originally discussed in Ref.~\cite{OrusTopologicalGE}, where it was shown that a transition occurs at $g_c \approx 0.56$ by calculating the topological contribution of the geometric entanglement.  The wavefunction with the string tension can be represented by the following tensor product state 
\begin{align}
|\Psi \rangle = \sum_{s_i} tTr (A^{s_1} B^{s_2}.... )   |s_1,s_2,....\rangle, 
\end{align}
where two tensors $A^s_{\alpha \beta \gamma \delta}$ and  $B^s_{\alpha \beta \gamma \delta}$ (on two respective sublattices of the square lattice) with virtual indices ${\alpha,\beta,\gamma,\delta}=\{0,1\}$ and the physical index $s=\{0,1\}$.
The non-zero coefficients of tensor are given by 
\begin{align}
&A^0_{0000}=1+g  \qquad   A^0_{1100}=1   \notag \\
&A^1_{1111}=1+g  \qquad  A^1_{0011}=1  \notag \\
&B^0_{0000}=1+g  \qquad   B^0_{1001}=1   \notag \\
&B^1_{1111}=1+g  \qquad  B^1_{0110}=1.
\end{align}

The calculation of the modular matrices can be performed analogously and  the gauge transformation in this case is given by 
$X=\bigl(\begin{smallmatrix}
0&1 \\ 1 &0
\end{smallmatrix} \bigr)$ acting on each bond index.
We obtain the modular matrices as $g$ varies, and the trace of these matrices  is shown in Fig.~\ref{fig:Z2string}.
At $0<g \leq 0.56 $ we have non-trivial modular matrices (see Eq.~\ref{nontri_oZ2}) with $tr(S)=tr(T)=2$.
At $g >0.56 $, we have trivial modular matrices [see Eq.~(\ref{trivial_Z2})] with $tr(S)=tr(T)=1$.

\section{ The $\mathbb{Z}_2$ CZX model with string tension} \label{app:CZX}

Consider a 2D square lattice with four spin-$1/2$ particles per site and the four neighboring spins form a plaquette-like entangled state $|\psi_{p}\rangle =  |0000\rangle + |1111\rangle$.    
This state, denoted by $|\Psi_{CZX}\rangle$ is the ground state of  the CZX model~\cite{Chen_2DCZX_2011} and is invariant under the special onsite $\mathbb{Z}_2$ symmetry, which is 
 generated by $U_{czx} = U_X U_{CZ}$, 
where  $U_X = X_1 \otimes X_2 \otimes X_3 \otimes X_4 $, and $X_i$ is a Pauli matrix 
$\bigl(\begin{smallmatrix}
0&1 \\ 1 &0
\end{smallmatrix} \bigr)$ on the spin $i$  and  $U_{CZ} = CZ_{12} \otimes CZ_{23} \otimes CZ_{34} \otimes CZ_{41} $,
where
\begin{equation}
 CZ_{ij}=|00\rangle\langle 00| +|01\rangle\langle 01|+|10\rangle\langle 10|-|11\rangle\langle 11|
 \end{equation}
 is the controlled-$Z$ operator on two spins labeled $i$ and $j$.

We then consider a one-parameter deformation  
$D(g)=diag(1,g,g,g,g,g^2,g,g,g,g,g^2,g,g,g,g,1) $ on each site, which is diagonal in the four-qubit computational basis.  Such deformation gives rise to a one-parameter family of states invariant under $U_{czx}$:
\begin{align}
|\Psi\rangle  \propto  D(g)^{\otimes N} |\Psi_{CZX}\rangle,
\end{align}
which can be represented by a local tensor $A[s_1,s_2,s_3,s_4]$ as shown in Fig.~\ref{fig:square_local tensor}, as follows:
\begin{align}
& A[0,0,0,0] = A[1,1,1,1] = 1\notag \\
& A[0,0,1,1] = A[1,1,0,0] = A[1,0,0,1] = A[0,1,1,0] =  g^2  \notag \\
& A[0,1,1,1] = A[1,0,0,0] = A[0,0,1,0] = A[1,1,0,1] = g^2 \notag \\
& A[0,0,0,1] = A[1,1,1,0] = g^2  \notag \\
& A[0,1,0,0] = A[1,0,1,1] = g^2  \notag \\
& A[0,1,0,1] = A[1,0,1,0] = g^4.
\label{CZX_deform_tensor}
\end{align}

\begin{figure}[t]
\center{\epsfig{figure=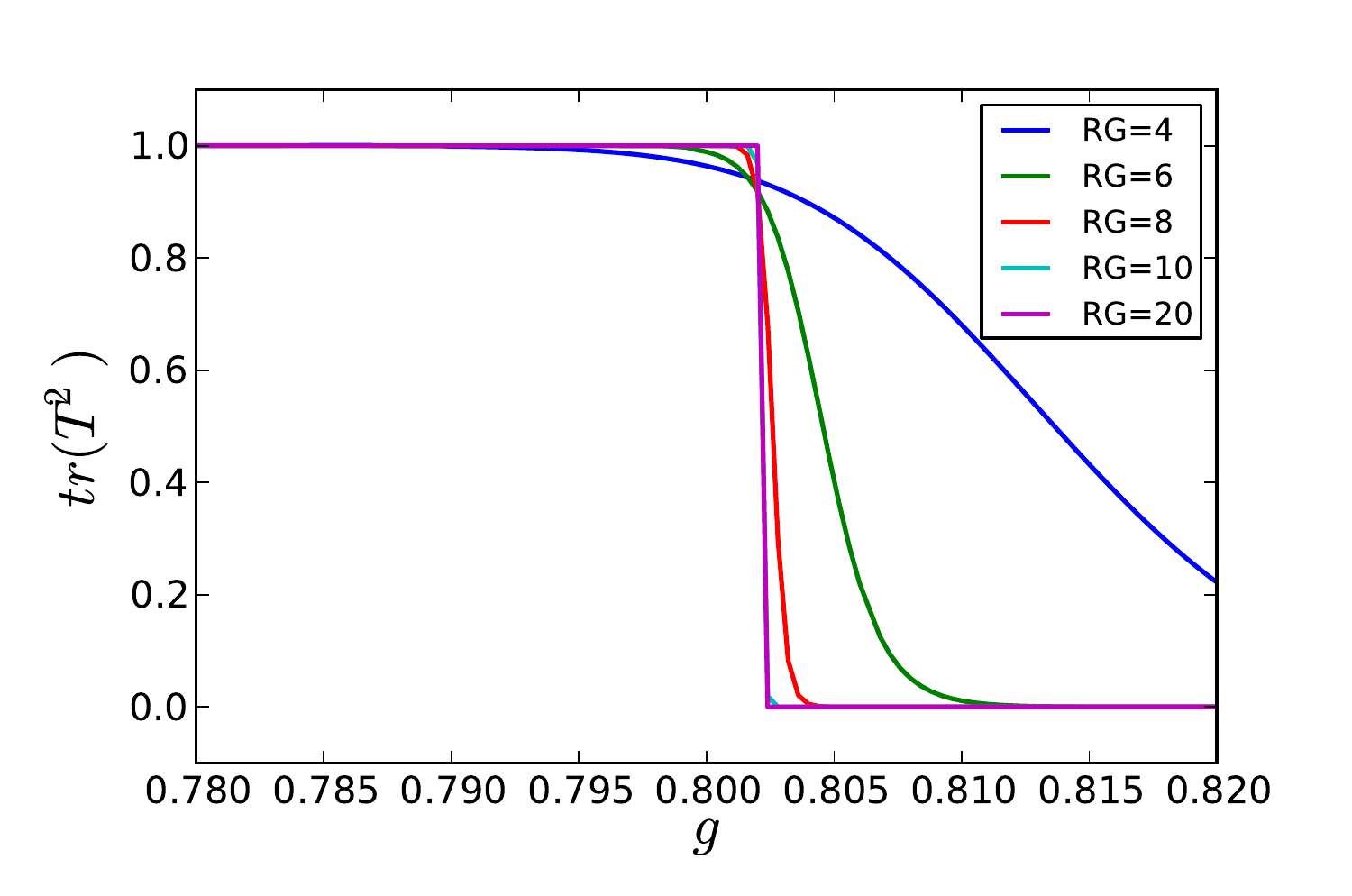,angle=0,width=8cm}}
\caption{ The deformed CZX model: the trace of $T^2$ as a function of parameter $g$ with $D_{cut}=16$ as increasing RG steps. The critical point is at $g_c \approx 0.802$. }
  \label{fig:CZX_deform}
\end{figure}

At $g=1$, this is the ground state of CZX model and the corresponding state has non-trivial $\mathbb{Z}_2$  SPT order. 
One may noticed that this state is also the ground state of the Hamiltonian [see Eq.~(\ref{eq:z2spt_h0})] with trivial $\mathbb{Z}_2$  SPT order.
The two models are different in their symmetry actions. 
In the previous trivial $\mathbb{Z}_2$ model,  the $\mathbb{Z}_2$  symmetry action is generated by operator $X$. 
By comparison, the symmetry of CZX model is more complicated. 
At $g=0$, the tensor represents a cat state $  |00...00\rangle + |11...11\rangle$ and such a global superposition represents the spontaneous symmetry breaking (i.e., unstable under an infinitesimal symmetry breaking field). 
At some critical point in $g_c$, the state must undergo a phase transition.

One way to detect the transition is to apply our algorithm. 
The MPO of deformed CZX model is independent of $g$ and is given by $X \otimes X \alpha$, with $\alpha = | 00\rangle \langle 00 | +  | 01\rangle \langle 01 | + | 10\rangle \langle 10 | - | 11\rangle \langle 11 |$.
We find that, for $g>g_c$, the state possesses nontrivial $\mathbb{Z}_2$  SPT order, since we obtain the nontrivial matrix $T^2$; see Eq.~(\ref{T2matrixk1}).  
While $g<g_c$, we obtain the same $T^2$ matrix as in Eq.~(\ref{T2matrix_tri}), and this region is a spontaneous symmetry-breaking phase.
We determine $g_c$ to lie between $0.802$ and $0.8024$ as shown in Fig.~\ref{fig:CZX_deform}. 

As a confirmation of the transition point, we find that 
the norm square of the wavefunction, given by the contraction of the double tensor over the lattice,  is just two copies of the partition function of the classical Ising model on the square lattice. 
From this  we then obtain the relation, $g^4= {(e^{\beta}-1)}/{(e^{\beta}+1)}$, between the parameter $g$ and the inverse temperature $\beta$, and the transition point is thus located at $g_c=0.802243$, with which  
our result agrees very well.
Our calculation for the quantum model confirms that there is  a quantum phase transition   between a nontrivial SPT phase and a symmetry-breaking phase.

%\bibliographystyle{apsrev_nurl}	
%\bibliography{bibs}	

%merlin.mbs apsrev4-1.bst 2010-07-25 4.21a (PWD, AO, DPC) hacked
%Control: key (0)
%Control: author (8) initials jnrlst
%Control: editor formatted (1) identically to author
%Control: production of article title (-1) disabled
%Control: page (0) single
%Control: year (1) truncated
%Control: production of eprint (0) enabled
%

 \end{document}